\documentclass[10pt]{article}
\usepackage{amsmath,amssymb}
\usepackage{newtxtext,newtxmath,bm}
\usepackage{graphics,graphicx,epsfig}
\usepackage{natbib}
\usepackage[hmargin=1in,vmargin=1in]{geometry}
\numberwithin{equation}{section}

\begin{document}

%
\title{Anti-plane segregation and diffusion in dense, bidisperse granular shear flow}
\author{Harkirat Singh and David L. Henann\footnotemark[2]\\
School of Engineering, Brown University, Providence, RI 02912, USA}
\renewcommand*{\thefootnote}{\fnsymbol{footnote}}
\footnotetext[2]{Email address for correspondence: david\_henann@brown.edu}
\renewcommand*{\thefootnote}{\arabic{footnote}}
\date{}
\maketitle
\begin{abstract}
Many dense granular systems are non-monodisperse, consisting of particles of different sizes, and will segregate based on size during flow. This phenomenon is an important aspect of many industrial and geophysical processes, necessitating predictive continuum models. This paper systematically studies a key aspect of the three-dimensional nature of segregation and diffusion in flowing, dense, bidisperse granular mixtures---namely, segregation and diffusion acting along the direction perpendicular to the plane of shearing, which we refer to as the anti-plane modes of segregation and diffusion. To this end, we consider discrete-element method (DEM) simulations of flows of dense, bidisperse mixtures of frictional spheres in an idealized configuration that isolates anti-plane segregation and diffusion. We find that previously-developed constitutive equations, calibrated to DEM simulation results from flows in which both the segregation and diffusion processes occur within the plane of shearing, do not capture aspects of the anti-plane segregation dynamics. Accordingly, we utilize DEM simulation results to inform and calibrate constitutive equations for the segregation and diffusion fluxes in their anti-plane modes. Predictions of the resulting continuum model for the anti-plane segregation dynamics are tested against additional DEM simulation results across different cases, while parameters such as the shear strain rate and mixture composition are varied, and we find that the calibrated model predictions match well with the DEM simulation results. Finally, we suggest a strategy for generalizing the constitutive forms for the segregation and diffusion fluxes to obtain three-dimensional constitutive equations that account for both the in-plane and anti-plane modes of the segregation and diffusion processes.
\end{abstract}
%

\section{Introduction}\label{sec:intro}
Dense granular mixtures consisting of particles of different sizes tend to segregate, or demix, based on size during shear flow. This phenomena is an important aspect of geophysical flows as well as industrial processes involving blending of granular mixtures. Continuum models capable of predicting the evolution of the spatially inhomogeneous composition of a granular mixture are valuable in understanding or designing processes involving size segregation. Consequently, substantial effort has gone into developing continuum constitutive equations for segregation and diffusion fluxes in flowing, dense granular mixtures based on both simulated and experimental granular systems \citep[e.g.,][]{gray05,gray06,fan11b,marks2012,hill2014,fan14,gajjar2014asymmetric,tunuguntla2014mixture,jones2018asymmetric,fry2018effect,trewhela2021experimental,liu2023coupled,singh_liu_arxiv}, primarily focusing on bidisperse mixtures made up of grains of two different sizes.

The vast majority of works on continuum modeling of size segregation in dense, bidisperse granular flows---including all of those cited in the previous paragraph---have dealt with two-dimensional flow and segregation scenarios. More precisely, denoting the plane of shearing as the $x$-$y$-plane, the components of the velocity field are of the form $v_x(x,y)$, $v_y(x,y)$, and $v_z=0$, and the vector describing the relative flux of large and small grains---which accounts for both the segregation and diffusion processes---lies entirely within the $x$-$y$-plane. Throughout this paper, we refer to the segregation and diffusion processes in this scenario as their ``in-plane'' modes. Common examples from the literature include flow down an inclined plane \citep[e.g.,][]{gray05,gray06,marks2012,gajjar2014asymmetric,tunuguntla2014mixture,singh_liu_arxiv}, two-dimensional bounded heap flow \citep[e.g.,][]{fan14,jones2018asymmetric,schlick2015}, and flow in a rotating drum \citep[e.g.,][]{hill2014,liu2019modeling,barker2021coupling,yang2021continuum}. However, segregation and diffusion in many bidisperse shear flows---e.g., split-bottom flow \citep{hill08,fan10}---are not in their in-plane modes. It has not been established whether flux constitutive equations proposed and calibrated based on measurements from discrete simulations or experiments involving the in-plane modes of segregation and diffusion remain predictive in more general, three-dimensional scenarios, which serves as the motivation for this work.

The purpose of this paper is to systematically study the three-dimensional nature of the continuum constitutive equations for the segregation and diffusion fluxes in flowing, dense, bidisperse granular mixtures, using discrete-element method (DEM) simulations of frictional spheres. In particular, we consider an idealized flow configuration, in which a semi-infinite layer of dense granular material is sheared within the $x$-$y$-plane but gravity and composition gradients act along the $z$-direction, so that the segregation and diffusion fluxes act along the direction perpendicular to the plane of shear. Throughout this paper, we refer to the segregation and diffusion processes in this scenario as their ``anti-plane'' modes. We then test whether a continuum model for the in-plane modes of segregation and diffusion in a dense, bidisperse mixture is capable of capturing the evolution of the spatially inhomogeneous composition of the granular mixture in this anti-plane scenario. We focus on the continuum model proposed in our prior work \citep{singh_liu_arxiv}, which can predict the dynamics of segregation across different cases of flow down an inclined plane as well as planar shear flow with gravity. Observations of size segregation in both of these flow configurations are examples of the in-plane mode of segregation. We find that the in-plane segregation model fails to capture several features of the segregation dynamics in the anti-plane mode---namely, the rate of diffusion and segregation at sufficiently low shear strain rates (quantified through the inertial number) and the scaling of the segregation flux with pressure. To remedy these points, we first isolate the anti-plane mode of diffusion by removing segregation due to gravity to motivate a modified constitutive equation for the diffusion flux. Second, we return to the anti-plane mode of segregation driven by a gravitational pressure gradient perpendicular to the plane of shearing to inform a modified constitutive equation for the segregation flux. When combined, the modified continuum model is capable of capturing the anti-plane segregation dynamics across different combinations of shear strain rate, mixture composition, and layer thickness.

This paper is organized as follows. We first discuss the continuum framework used to describe dense, bidisperse granular systems in Section~\ref{sec:conti_frame}. Then, we introduce the configuration for anti-plane segregation in simple shear flow in Section~\ref{sec:antiplaneseg} and illustrate the deficiencies of the planar model of \citet{singh_liu_arxiv}. Next, based on DEM simulation data, we propose continuum constitutive equations for the anti-plane modes of the diffusion and segregation fluxes in Sections~\ref{sec:diffusion} and \ref{sec:segregation}, respectively. The proposed continuum model is solved to predict the transient evolution of the concentration field in anti-plane segregation in simple shear flow, and continuum model predictions are tested against DEM simulation results in Section~\ref{sec:validation}. Revisiting anti-plane diffusion, in Section~\ref{sec:MSD_anti_plane_diffusion}, we characterize the diffusion flux through the mean square displacement (MSD) of the granular mixture and contrast the result with the diffusion flux characterized through the dynamics of mixing of an initially segregated granular mixture from Section~\ref{sec:diffusion}. Lastly, we describe a strategy for generalizing the constitutive equations for the segregation and diffusion fluxes to obtain three-dimensional constitutive forms that are capable of switching between the respective in-plane and anti-plane modes in Section \ref{sec:3D_constitutive_form}. We conclude the paper with some closing remarks in Section \ref{sec:conclusion}. 

\section{Continuum framework}\label{sec:conti_frame}
In this section, we discuss the continuum framework used to describe segregation and diffusion in dense, bidisperse granular mixtures. Regarding notation, we use component notation, in which the components of vectors, $\boldsymbol{v}$, and tensors, $\boldsymbol{\sigma}$, relative to a set of Cartesian basis vectors $\{\boldsymbol{e}_i|i=1,2,3\}$ are denoted by $v_i$ and $\sigma_{ij}$, respectively, and the Einstein summation convention is employed.

We consider bidisperse granular mixtures consisting of grains of two different sizes---large grains with diameter $d^{\rm l}$ and small grains with diameter $d^{\rm s}$. All grains are made of same material with mass density $\rho_{\rm s}$, thereby eliminating density-based segregation and isolating size-based segregation. To describe the dynamics of segregation in dense, bidisperse granular systems at the continuum level, we utilize a mixture-theory approach, which is common in the literature \citep[e.g.,][]{gray2018,umbanhowar2019modeling}. The solid volume fractions of each species are denoted as $\phi^{\rm l}$ and $\phi^{\rm s}$ for the large and small grains, respectively. For the dense shear flows considered in the present work, volume changes are minimal, and any volume change at flow initiation occurs over a much shorter time scale than the process of segregation. Accordingly, we adopt the idealization that the total solid volume fraction $\phi = \phi^{\rm l} + \phi^{\rm s}$ is constant and, based on observations from DEM simulations, approximately equal to $\phi=0.6$ for dense spheres. The concentrations of each species are $c^{\rm l} = \phi^{\rm l}/\phi$ and $c^{\rm s} = \phi^{\rm s}/\phi$, so that $c^{\rm l}  + c^{\rm s} =1$. The average grain size of the mixture is then defined as $\bar{d} = c^{\rm l} d^{\rm l} + c^{\rm s} d^{\rm s}$. 

The mixture velocity $v_i$ is defined through the partial velocities of each species, $v_i^{\rm l}$ and $v_i^{\rm s}$, as $v_i = c^{\rm l} v^{\rm l}_i + c^{\rm s} v^{\rm s}_i$. Then, the strain-rate tensor for the mixture is $D_{ij} = (1/2)\left(\partial v_i/ \partial x_j  + \partial v_j/\partial x_i\right)$, and the equivalent shear strain rate is $\dot{\gamma} = (2 D_{ij} D_{ij})^{1/2}$. Due to the constant-volume idealization, the mixture velocity is divergence-free, $\partial v_i/\partial x_i = 0$, and the mixture strain-rate tensor is deviatoric, $D_{kk}=0$. The relevant stress-related fields for the mixture are the symmetric Cauchy stress tensor $\sigma_{ij}=\sigma_{ji}$ and the pressure $P=-(1/3)\sigma_{kk}$. The inertial number for a bidisperse system is defined in terms of the average grain size as $I = \dot\gamma\bar{d}\sqrt{\rho_{\rm s}/P}$ \citep{rognon07,tripathi11}.

The relative volume flux of species $\nu ={\rm l}, {\rm s}$ is defined as $w^{\nu}_i  = c^{\nu}(v^{\nu}_i - v_i)$, so that $w^{\rm l} + w^{\rm s} = 0$. We utilize $c^{\rm l}$ as the field variable that describes the segregation dynamics, and under the constant-volume idealization, the mass conservation equation for the large grains governs the evolution of $c^{\rm l}$ as follows:  
\begin{equation}\label{eq:cons_mass}
\frac{D c^{\rm l}}{D t} + \frac{\partial w^{\rm l}_i}{\partial x_i} = 0, 
\end{equation}
where $D(\bullet)/Dt$ is the material time derivative. To close the segregation model, a constitutive equation for the relative volume flux of large particles $w^{\rm l}_i$ is needed. We take the flux to be comprised of two contributions:
\begin{equation}\label{eq:fluxdecomp}
    w^{\rm l}_i = w^{\rm diff}_i + w^{\rm P}_i,
\end{equation}
where $w^{\rm diff}_i$ is the diffusion flux that acts to remix the species and $w^{\rm P}_i$ is the pressure-gradient-driven segregation flux that acts to demix the species. In this paper, we focus on simple shear flows, in which the equivalent shear strain rate $\dot\gamma$ is spatially uniform, so the shear-strain-rate-gradient-driven segregation of \citet{liu2023coupled} does not appear in \eqref{eq:fluxdecomp}.

\section{Anti-plane segregation in simple shear flow}\label{sec:antiplaneseg}
\subsection{Discrete-element method simulations}
\begin{figure}[!t]
\centering
\includegraphics{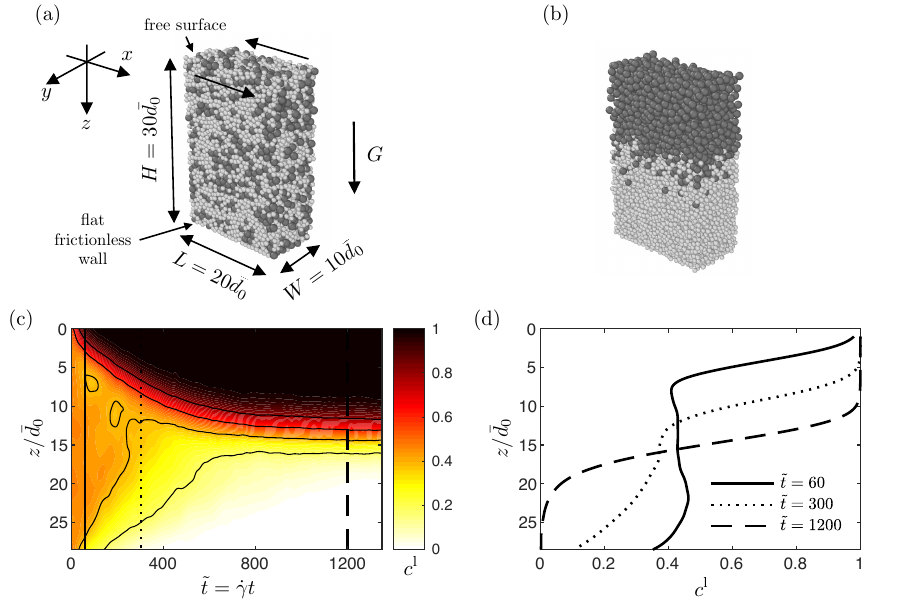} 
\caption{Representative base case for anti-plane segregation in simple shear flow. (a) Initial well-mixed configuration for three-dimensional DEM simulation of bidisperese spheres. (b) Segregated state after a simulation time of $\tilde{t} = \dot\gamma t = 1200$. (c) Spatiotemporal evolution of the $c^{\rm l}$ field. (d) Spatial profiles of the $c^{\rm l}$ field at three different time instants ($\tilde{t} = 60$, 300, and 1200) as indicated by the three vertical lines in (c).}\label{fig:aplane_seg}
\end{figure}

In this section, we examine the anti-plane mode of segregation by considering simple shear flow of a layer of a dense, bidisperse granular mixture within the $x$-$y$-plane. The layer has a finite thickness $H$ along the $z$-direction, which is perpendicular to the plane of shearing, and gravity $G$ acts along the $z$-direction. We consider simple shear flows of dense, bidisperse spheres with an inter-particle friction coefficient of $\mu_{\rm surf}=0.4$ using discrete-element method (DEM) simulations. Further details of the simulated granular system are given in Appendix~\ref{app:appendix_A}. The DEM setup for the case of a well-mixed system of bidisperse spheres and $H=30\bar{d}_0$ is shown in Fig.~\ref{fig:aplane_seg}(a), where $\bar{d}_0$ is the system-wide average grain size. Shearing in the $x$-$y$ plane is achieved through Lees-Edwards boundary conditions \citep{lees1972computer}. The dimensions of the DEM simulation domain are taken to be $L= 20\bar{d}_0$ along the $x$-direction and $W=10 \bar{d}_0$ along the $y$-direction. The bottom boundary is a flat and frictionless wall that does not impart any shear traction within its plane, and the top surface is traction-free. This setup achieves a homogeneous state of strain rate, in which the only non-zero components of the strain-rate tensor are $D_{xy} = D_{yx} = \dot\gamma/2$, and eliminates segregation due to strain-rate gradients \citep[e.g.,][]{liu2023coupled}, isolating pressure-gradient-driven segregation in its anti-plane mode. As discussed in Section~\ref{sec:intro}, we use the term ``anti-plane'' to designate that pressure-gradient-driven size segregation acts along the direction (the $z$-direction) perpendicular to the plane of shearing (the $x$-$y$ plane), which contrasts with the ``in-plane'' mode examined in our prior work \citep{singh_liu_arxiv}, where segregation acts along a direction that is within the plane of shearing.

The important dimensionless parameters that characterize anti-plane segregation in simple shear flow are (1) the dimensionless layer thickness $H/\bar{d}_0$, (2) the dimensionless strain rate $\dot{\gamma}\sqrt{\bar{d}_0/G}$, (3) the initial large-grain concentration field $c^{\rm l}_0(z)$, and (4) the bidisperse grain-size ratio $d^{\rm l}/d^{\rm s}$. DEM simulation results for a representative base case corresponding to the parameter set $\{H/\bar{d}_0 = 30, \dot{\gamma}\sqrt{\bar{d}_0/G} = 0.06, c^{\rm l}_0 = 0.50, d^{\rm l}/d^{\rm s} = 1.5  \}$ are summarized in Fig~\ref{fig:aplane_seg}. The segregated state after a simulation time of $\tilde{t} = \dot\gamma t = 1200$ is shown in Fig.~\ref{fig:aplane_seg}(b), showing that the large grains (dark gray) indeed segregate to the top free surface, leaving small grains (light gray) on the bottom of the layer. To further quantify the process of segregation, we coarse-grain the large-grain concentration field $c^{\rm l}$ and plot spatiotemporal contours of the evolution of $c^{\rm l}$ in Fig.~\ref{fig:aplane_seg}(c). (Coarse-graining procedures are described in Appendix~\ref{app:appendix_A}.) The growing width of the dark region indicates the accumulation of large grains at the top of the layer. Finally, in Fig.~\ref{fig:aplane_seg}(d), we plot spatial profiles of the $c^{\rm l}$ field along the $z$-direction at three snapshots in time, associated with short, moderate, and long times ($\tilde{t} = 60, 300, 1200$) with respect to the process of segregation, indicated by the vertical lines in Fig.~\ref{fig:aplane_seg}(c).

\subsection{Continuum modeling based on in-plane constitutive equations}\label{sec:inplane}
Next, we test the segregation model described in \citet{singh_liu_arxiv} for the in-plane modes of segregation and diffusion in a bidisperse mixture in the context of anti-plane segregation and diffusion. First, we briefly recap the constitutive equation for the relative volume flux $w^{\rm l}_i$, which comprises contributions due to diffusion and segregation. The diffusion flux $w^{\rm diff}_i$ acts to remix the two species and is taken to be driven by concentration gradients as $w^{\rm diff}_i = - D(\partial c^{\rm l}/\partial x_i)$, where $D$ is the binary diffusion coefficient which scales with the mean grain size $\bar{d}$ and the shear strain rate $\dot\gamma$ as $D\sim \bar{d}^2\dot\gamma$ \citep{utter2004,fry2019diffusion,bancroft2021drag,liu2023coupled}. Accordingly, the diffusion flux is
\begin{equation}\label{eq:diff_eqn}
w^{\rm diff}_i = - C_{\rm diff}^{\rm in}\bar{d}^2\dot\gamma \dfrac{\partial c^{\rm l}}{\partial x_i},
\end{equation}
where $C_{\rm diff}^{\rm in}$ is a dimensionless, constant material parameter that characterizes the in-plane diffusion flux. Regarding the segregation flux, in the present work, we consider homogeneous shear flows, in which the shear strain rate $\dot\gamma$ is spatially uniform. Therefore, shear-strain-rate gradients \citep[e.g.,][]{liu2023coupled} do not drive the segregation process, and we only consider segregation driven by pressure gradients \citep[e.g.,][]{singh_liu_arxiv} and adopt the following constitutive equation for the segregation flux $w^{\rm P}_i$: 
\begin{equation}\label{eq:segP_eqn}
w^{\rm P}_i = - C^{\rm P,in}_{\rm seg}\dfrac{\bar{d}^2\dot\gamma}{P}c^{\rm l}(1 - c^{\rm l})(1 - \alpha + \alpha c^{\rm l})\dfrac{\partial P}{\partial x_i},
\end{equation}
where $C^{\rm P,in}_{\rm seg}$ and $\alpha$ are dimensionless, constant parameters that characterize the in-plane segregation flux. The scaling of the pre-factor in \eqref{eq:segP_eqn} with $\bar{d}^2\dot\gamma/P$ follows prior works in the literature \citep{trewhela2021experimental,barker2021coupling,singh_liu_arxiv} and ensures dimensional consistency. Moreover, this scaling captures the requirement that the segregation flux is zero when there is no flow $(\dot\gamma = 0)$ as well as the attenuation of segregation with increasing pressure \citep{golick2009mixing,fry2018effect}. The dependence of the pre-factor in \eqref{eq:segP_eqn} on $c^{\rm l}$ through the function $f(c^{\rm l}) = c^{\rm l}(1 - c^{\rm l})(1 - \alpha + \alpha c^{\rm l})$ follows from \citet{gajjar2014asymmetric} and captures the asymmetric dependence of the flux on $c^{\rm l}$, quantified by the dimensionless parameter $\alpha \in [0, 1]$. For $\alpha=0$, the dependence of $f(c^{\rm l})$ is symmetric about its maximum at $c^{\rm l}=0.5$, but for $\alpha\in(0,1]$, the maximum of $f(c^{\rm l})$ skews from $c^{\rm l}=0.5$ towards $c^{\rm l}=1$, while maintaining that segregation ceases when the bidisperse mixture becomes either all large ($c^{\rm l}=1$) or all small ($c^{\rm l}=0$) grains. While it is well established in the literature that the pressure-gradient-driven segregation flux depends on the grain-size ratio $d^{\rm l}/d^{\rm s}$ \citep[e.g.,][]{schlick2015,jones2018asymmetric,trewhela2021experimental}, we focus on a single grain-size ratio of $d^{\rm l}/d^{\rm s}=1.5$ throughout this work, and therefore, this dependence does not appear in \eqref{eq:segP_eqn}. The relative volume flux $w^{\rm l}_i$ is then taken to be the sum of \eqref{eq:diff_eqn} and \eqref{eq:segP_eqn}: $w^{\rm l}_i = w^{\rm diff}_i + w^{\rm P}_i$, where the material parameters associated with the in-plane segregation model for a given $d^{\rm l}/d^{\rm s}$ are $\{C_{\rm diff}^{\rm in},C^{\rm P,in}_{\rm seg},\alpha\}$. 

Utilizing the flux constitutive equations \eqref{eq:diff_eqn} and \eqref{eq:segP_eqn} in the balance of mass equation \eqref{eq:cons_mass}, we obtain a governing equation for the $c^{\rm l}$ field in the absence of shear-strain-rate gradients. In the context of anti-plane segregation in simple shear flow (Fig.~\ref{fig:aplane_seg}(a)), in which the concentration and pressure fields only vary along the $z$-direction, the large-grain concentration field $c^{\rm l}(z,t)$ is governed by the following partial differential equation (PDE): 
\begin{equation}\label{eq:seg_evol_inplane}
\frac{\partial c^{\rm l}}{\partial t} + \frac{\partial }{\partial z} \left( -C_{\rm diff}^{\rm in} \bar{d}^2 \dot{\gamma} \frac{\partial c^{\rm l}}{\partial z}  - {C}^{\rm P,in}_{\rm seg} \frac{\bar{d}^2 \dot{\gamma}}{P} c^{\rm l}(1- c^{\rm l})(1- \alpha + \alpha c^{\rm l}) \frac{\partial P}{\partial z} \right) =0.
\end{equation}
In this configuration, the strain rate $\dot\gamma$ is prescribed and is spatially and temporally uniform. Moreover, the stress field does not vary in time, and the out-of-plane normal stress component $\sigma_{zz}(z)$ may be determined from a static force balance along the $z$-direction to be $\sigma_{zz}(z) = -\phi\rho_{\rm s} G z$. The in-plane normal stresses $\sigma_{xx}(z)$ and $\sigma_{yy}(z)$ are equal and, in coarse-grained DEM data, are observed to also be linearly proportional to $z$; however, due to the normal stress differences that arise in dense flows of spheres \citep{depken2007,srivastava2021viscometric}, the magnitude of the slope of this linear dependence is observed to be slightly higher than $\phi\rho_{\rm s} G$. As a result, the pressure field $P(z)$ varies linearly in $z$ but with a constant slope $\partial P/\partial z$ that is slightly higher than the nominal value of $\phi\rho_{\rm s} G$. To control for this effect, when obtaining continuum solutions, we utilize the value of the slope of the pressure field $\partial P/\partial z$ obtained from the coarse-grained stress fields in the DEM data, rather than the nominal value of $\phi\rho_{\rm s} G$. Furthermore, to avoid a singularity in the pressure-gradient-driven segregation flux at the free surface where $z=0$, we add a small constant to the pressure field corresponding to the weight of a layer of $(1/4)\bar{d}_0$ thickness, i.e., $(1/4)(\partial P/\partial z)\bar{d}_0$. We have verified that continuum model predictions are insensitive to the exact choice of this constant, so long as it is sufficiently small. Therefore, with $P(z) = (\partial P/\partial z)z + (1/4)(\partial P/\partial z)\bar{d}_0$, we have that $(1/P)\partial P/\partial z = 1/(z + (1/4)\bar{d}_0)$ in the last term of \eqref{eq:seg_evol_inplane}. 

When accompanied by boundary and initial conditions, \eqref{eq:seg_evol_inplane} may then be used to obtain predictions for the transient evolution of the large-grain concentration field $c^{\rm l}(z,t)$ during anti-plane segregation in simple shear flow. Regarding boundary conditions, we impose no-flux boundary conditions at the top free surface as well as at the bottom frictionless wall, i.e., $w^{\rm l}_z = -{C}_{\rm diff}^{\rm in} \bar{d}^2 \dot{\gamma}  ({\partial c^{\rm l}}/{\partial z})  - {C}^{\rm P,in}_{\rm seg} ({\bar{d}^2 \dot{\gamma}}/{P}) c^{\rm l}(1- c^{\rm l})(1- \alpha + \alpha c^{\rm l}) ({\partial P}/{\partial z}) = 0$ at $z=0$ and $H$. For the initial condition for the large-grain concentration field $c^{\rm l}_0(z)=c^{\rm l}(z, t=0)$, we obtain the coarse-grained field from the initial DEM configuration at $t=0$ and use this field as the initial condition $c^{\rm l}_0(z)$ in the corresponding continuum simulation. This is done to control for spatial concentration fluctuations in the initial DEM state. Then, for a given case of anti-plane segregation in simple shear flow, described by the parameter set $\{H/\bar{d}_0, \dot{\gamma}\sqrt{\bar{d}_0/G}, c^{\rm l}_0, d^{\rm l}/d^{\rm s}\}$, we obtain numerical solutions using finite differences. Central differences are used to discretize the spatial derivatives in \eqref{eq:seg_evol_inplane}. The Euler method is used for temporal discretization, in which the spatial derivative of $c^{\rm l}$ appearing in the diffusion flux term in \eqref{eq:seg_evol_inplane} is treated implicitly for improved numerical stability while the pre-factors of both flux terms in \eqref{eq:seg_evol_inplane} are treated explicitly. We use sufficiently small spatial and temporal resolutions in the finite-difference scheme to ensure stable and accurate results. 

\begin{figure}[!t]
\centering
\includegraphics{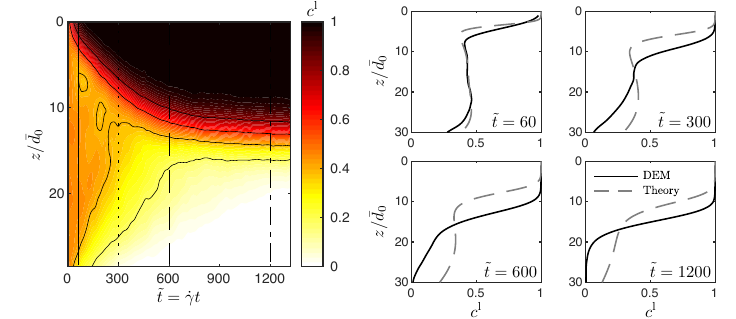} 
\caption{Comparisons of predictions of the in-plane continuum model of \citet{singh_liu_arxiv} with corresponding DEM simulation results for the transient evolution of the segregation dynamics for the base case of anti-plane segregation in simple shear flow $\{H/\bar{d}_0 = 30, \dot{\gamma}\sqrt{\bar{d}_0/G} = 0.06, c^{\rm l}_0 = 0.50, d^{\rm l}/d^{\rm s} = 1.5  \}$. On the left are spatiotemporal contours of the evolution of $c^{\rm l}$ measured in the DEM simulation, and on the right are comparisons of DEM simulation results (solid black lines) and continuum model predictions (dashed gray lines) for the $c^{\rm l}$ field at four different time instants during the segregation process: $\tilde{t} = 60$, 300, 600, and 1200 in the sequence of top left, top right, bottom left, bottom right and indicated by the vertical lines on the contour plot.}\label{fig:inplane_model}
\end{figure}

We compare predictions of the in-plane-based continuum model \eqref{eq:seg_evol_inplane} against DEM data for the representative base case of anti-plane segregation in simple shear flow from Fig.~\ref{fig:aplane_seg}, $\{H/\bar{d}_0 = 30, \dot{\gamma}\sqrt{\bar{d}_0/G} = 0.06, c^{\rm l}_0 = 0.50, d^{\rm l}/d^{\rm s} = 1.5\}$, using the in-plane parameters from \citet{singh_liu_arxiv}, i.e., $\{{C}_{\rm diff}^{\rm in}=0.045,{C}^{\rm P,in}_{\rm seg}=0.34,\alpha=0.4\}$. Comparisons of the continuum model predictions (dashed gray lines) with the DEM simulations (solid black lines) are summarized in Fig.~\ref{fig:inplane_model} at four different time instants during the segregation process: $\tilde{t} = 60$, 300, 600, and 1200. The in-plane model falls short in capturing the transient evolution of the segregation dynamics in the anti-plane mode of segregation. First, predictions of the segregation dynamics lag behind the DEM simulation results in time, and second, a sharp transition zone near the top is predicted by the continuum model at early times $(\tilde{t}\lesssim 300)$, which is contrary to DEM simulation observations that show a comparatively wider transition zone. The main objective of this paper to adapt the constitutive equations for diffusion and pressure-gradient-driven segregation to capture segregation dynamics in the anti-plane mode. To do so, we examine the anti-plane mode of diffusion in Section~\ref{sec:diffusion} and the anti-plane mode of pressure-gradient-driven segregation in Section~\ref{sec:segregation}.

\section{Constitutive equations for anti-plane diffusion and segregation}\label{sec:antiplanefluxes}
\subsection{Anti-plane diffusion flux} \label{sec:diffusion}
\begin{figure}[!t]
\centering
\includegraphics{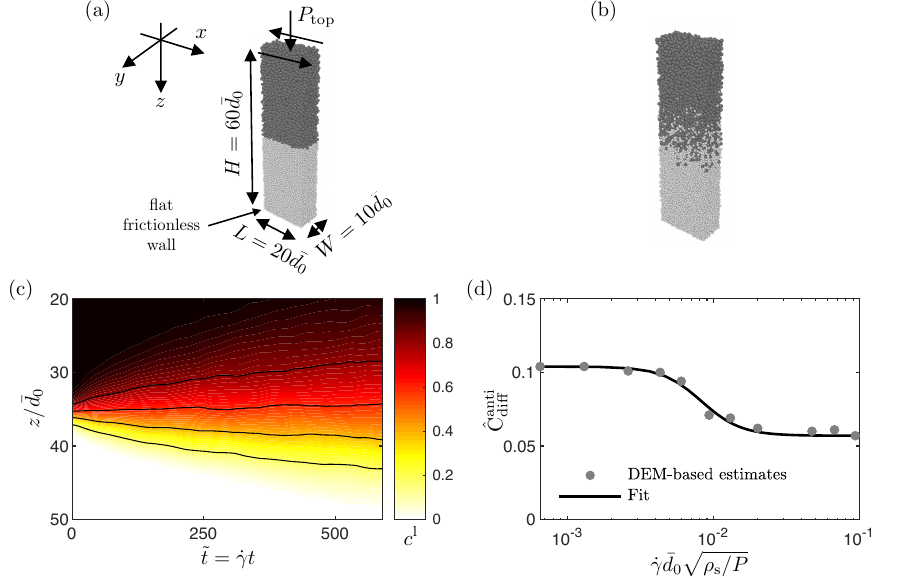} 
\caption{Representation of the anti-plane mode of diffusive mixing in simple shear flow. (a) Initially segregated configuration for three-dimensional DEM simulation of bidisperse spheres with $\sim 20~000$ grains. (b) Intermediate remixed state after shearing for a simulation time of $\tilde{t} = \dot\gamma t=590$. (c) Spatiotemporal evolution of the large-grain concentration field for a nominal inertial number of $\dot\gamma\bar{d}_0\sqrt{\rho_{\rm s}/P}=0.02$. (d) Estimated dependence of the anti-plane diffusion parameter $\hat{C}_{\rm diff}^{\rm anti}$ on the nominal inertial number $\dot\gamma \bar{d}_0\sqrt{\rho_{\rm s}/P}$. Symbols are estimates based on fitting the continuum model \eqref{eq:remix} to DEM data for different values of the nominal inertial number; the solid black curve represents the best fit of \eqref{eq:cd_anti} with $\{C_{0}=0.104, C_{\infty}=0.057, I_{\rm tr}=0.00744, m = 3.23\}$.}\label{fig:diffusion}
\end{figure}

In this section, we characterize the anti-plane mode of diffusion---i.e., when the diffusion process occurs along the direction perpendicular to the plane of shearing. To isolate the anti-plane diffusion flux, steady simple shear flow at a uniform strain rate $\dot\gamma$ is prescribed within the $x$-$y$-plane at a constant pressure $P$, while the system is initially segregated along the $z$-direction (the anti-plane direction). This is in contrast to the situation in Section~\ref{sec:antiplaneseg}, where the pressure field was not uniform but hydrostatic. A snapshot of the initial DEM configuration is shown in Fig.~\ref{fig:diffusion}(a), where large grains (dark gray) are on the top and small grains (light gray) are on the bottom, and the grain-size ratio is $d^{\rm l}/d^{\rm s}=1.5$. We employ Lees-Edwards boundary conditions within the $x$-$y$-plane to prescribe a uniform strain rate $\dot\gamma$, and the dimensions of the simulation domain along the $x$- and $y$-directions are $L=20 \bar{d}_0$ and $W=10 \bar{d}_0$, respectively. The size of the simulation domain along the $z$-direction is $H=60 \bar{d}_0$, and periodic boundary conditions are not employed along this direction.  Instead, the bottom boundary consists of a flat, frictionless wall, and a constant compressive normal stress is applied to the top through a mass of large grains (not pictured in Fig.~\ref{fig:diffusion}(a)) subjected to a gravitational body force along the $z$-direction. The grains in the region shown in Fig.~\ref{fig:diffusion}(a) are not subjected to any body forces. The result of these boundary conditions along the anti-plane direction is that the normal stress component $\sigma_{zz}$ is spatially and temporally constant and equal to $-P_{\rm top}$. Moreover, the normal stresses $\sigma_{xx}$ and $\sigma_{yy}$ are also uniform and compressive but of slightly higher magnitude due to normal stress differences that arise in dense flows of spheres. The spatially uniform pressure field is then $P = -(1/3)(\sigma_{xx}+\sigma_{yy}+\sigma_{zz})$ and is slightly greater than $P_{\rm top}$. 

Since the strain rate and pressure are uniform in this flow configuration, no size segregation takes place, and only diffusive mixing occurs. A DEM snapshot of an intermediate state after a simulation time of $\tilde{t} = \dot\gamma t=590$ is shown in Fig.~\ref{fig:diffusion}(b), illustrating that the initially sharp transition between large and small grains in Fig.~\ref{fig:diffusion}(a) becomes diffuse due to mixing. To further quantify the diffusion process, we coarse-grain the $c^{\rm l}$ field from the DEM simulations, and the spatiotemporal contours of the evolution of the $c^{\rm l}$ field are plotted in Fig.~\ref{fig:diffusion}(c) for the case of a nominal inertial number of $\dot\gamma\bar{d}_0\sqrt{\rho_{\rm s}/P}=0.02$, where $\bar{d}_0=(d^{\rm l} + d^{\rm s})/2$ and $P$ is the pressure obtained from the coarse-grained DEM results to control for the normal stress differences. We observe that the width of the transition region between large and small grains, which is initially sharp, grows in time as the diffusion process proceeds and the grains mix.
 
In the continuum model for anti-plane diffusion, we continue to take the diffusion flux $w^{\rm diff}_{i}$ to scale with the average grain size $\bar{d}$, the strain rate $\dot{\gamma}$, and the concentration gradient $\partial c^{\rm l}/\partial x_i$ as in \eqref{eq:diff_eqn}, so that
\begin{equation}\label{eq:flux_diff_anti}
w^{\rm diff}_i = -\hat{C}_{\rm diff}^{\rm anti} \bar{d}^2 \dot{\gamma} \dfrac{\partial c^{\rm l}}{\partial x_i}.
\end{equation}
The key difference from \eqref{eq:diff_eqn} is that we allow the dimensionless parameter associated with diffusion $\hat{C}_{\rm diff}^{\rm anti}$ to depend on additional dimensionless quantities---in particular, the inertial number $I=\dot\gamma\bar{d}\sqrt{\rho_{\rm s}/P}$. Plugging \eqref{eq:flux_diff_anti} in the mass conservation equation \eqref{eq:cons_mass} in the context of anti-plane diffusion in which the concentration field only varies along the $z$-direction, we obtain the following governing equation for the diffusion process:
\begin{equation}\label{eq:remix}
\frac{\partial c^{\rm l}}{\partial t} -   \frac{\partial}{\partial z} \left( \hat{C}_{\rm diff}^{\rm anti}  \bar{d}^2\dot{\gamma} \frac{\partial c^{\rm l}}{\partial z} \right) = 0. 
\end{equation}
To estimate the parameter $\hat{C}_{\rm diff}^{\rm anti}$ and its functional dependence on $I$, we perform DEM simulations of anti-plane diffusive mixing for 11 different values of the nominal inertial number $\dot\gamma \bar{d}_0\sqrt{\rho_{\rm s}/P}$ spanning from approximately 0.001 to 0.1. We then calculate predictions of \eqref{eq:remix} for the concentration field $c^{\rm l}(z,t)$ for different values of $\hat{C}_{\rm diff}^{\rm anti}$ over the range $\hat{C}_{\rm diff}^{\rm anti} \in (0.04,0.12)$. By minimizing the difference between the continuum model predictions and the coarse-grained concentration field from each DEM simulation, the best-fit value of $\hat{C}_{\rm diff}^{\rm anti}$ is determined for each value of the nominal inertial number $\dot\gamma \bar{d}_0\sqrt{\rho_{\rm s}/P}$. In the fitting process, we only consider sufficiently short dimensionless times $\tilde{t}$, corresponding to a transition width that remains substantially smaller than the layer thickness $H$, so that the diffusion process is unaffected by the boundaries. We note that this fitting process neglects the spatial variation in the inertial number field $I=\dot\gamma\bar{d}\sqrt{\rho_{\rm s}/P}$ that arises due to the spatial variation of $c^{\rm l}$ and hence $\bar{d}$. However, this simplification has a minimal effect on the estimates of $\hat{C}_{\rm diff}^{\rm anti}$ since, for a given nominal inertial number, the inertial number field varies by a factor of $d^{\rm l}/d^{\rm s}=1.5$ between the small grains at the bottom of the layer and the large grains on the top of the layer, while the inertial-number range-of-interest spans two orders of magnitude. Moreover, this simplification is only invoked during the fitting process, and, in all subsequent calculations, $\hat{C}_{\rm diff}^{\rm anti}$ is taken to be a function of the spatially varying inertial number field and not a global, nominal value of the inertial number. 

The best-fit values of $\hat{C}_{\rm diff}^{\rm anti}$ are plotted as a function of the nominal inertial number $\dot\gamma \bar{d}_0\sqrt{\rho_{\rm s}/P}$ as light gray circles in Fig.~\ref{fig:diffusion}(d). The estimated $\hat{C}_{\rm diff}^{\rm anti}$  plateaus to a lower value at high values of the nominal inertial number and to a higher value at low values of the nominal inertial number, and there is an intermediate range in which  $\hat{C}_{\rm diff}^{\rm anti}$ transitions from the higher plateau to the lower plateau. We use the following phenomenological functional form to fit the estimated values of the diffusion parameter over the range $0.001\lesssim I \lesssim 0.1$:
\begin{equation}\label{eq:cd_anti}
\hat{C}^{\rm anti}_{\rm diff}(I) = C_0 + \dfrac{C_\infty - C_0}{(I_{\rm tr}/I)^m + 1},
\end{equation}
where $C_0$ is the plateau value at low inertial numbers, $C_{\infty}$ is the plateau value at high inertial numbers, $I_{\rm tr}$ determines the position of the transition, and the exponent $m$ determines the width of the transition with greater values of $m$ corresponding to sharper transitions. Fitting \eqref{eq:cd_anti} to the DEM-based estimates in Fig.~\ref{fig:diffusion}(d) gives the parameter set associated with the anti-plane mode of diffusion for frictional spheres to be $\{C_{0}=0.104, C_{\infty}=0.057, I_{\rm tr}=0.00744, m = 3.23\}$. The fitted functional form is shown by the solid black line in Fig.~\ref{fig:diffusion}(d). In the semi-logarithmic plot of Fig.~\ref{fig:diffusion}(d), \eqref{eq:cd_anti} corresponds to hyperbolic-tangent-type behavior. 

The dependence of the anti-plane diffusion parameter $\hat{C}^{\rm anti}_{\rm diff}$ on the inertial number $I$ is in contrast to the in-plane diffusion parameter, which, in our experience \citep{liu2023coupled,singh_liu_arxiv}, may be idealized as independent of the inertial number, based on both mean square displacement DEM data and continuum-model predictions of the evolution of the $c^{\rm l}$ field. We note that the plateau value at high inertial numbers of $C_\infty=0.057$ is similar to the in-plane diffusion parameter $C_{\rm diff}^{\rm in}=0.045$ determined in \citet{liu2023coupled} and used in Section~\ref{sec:inplane}, so that diffusion is nearly isotropic for $I\gtrsim0.01$. However, we find that out-of-plane diffusion becomes comparatively stronger for low values of the inertial number with $\hat{C}^{\rm anti}_{\rm diff}$ increasing to a plateau value of $C_{0}=0.104$, indicating anisotropy in diffusion at low values of the inertial number. The dependence of $\hat{C}^{\rm anti}_{\rm diff}$ on $I$ is also in contrast to the mean square displacement DEM data of \citet{bancroft2021drag}, who found the diffusion coefficient along the anti-plane direction to be nearly constant with $\hat{C}^{\rm anti}_{\rm diff}\approx 0.04$ over a range of inertial numbers. We revisit this point in Section~\ref{sec:MSD_anti_plane_diffusion}. 
 
\subsection{Anti-plane segregation flux} \label{sec:segregation}
Next, we return to anti-plane segregation in simple shear flow as described in Section~\ref{sec:antiplaneseg} to characterize the anti-plane mode of segregation. We take the pressure-gradient-driven segregation flux to be given in the following form:
\begin{equation}\label{eq:seg_flux}
w^{\rm P}_i = -\hat{C}^{\rm P,anti}_{\rm seg} \frac{\bar{d}^2 \dot{\gamma}}{P} c^{\rm l}(1-c^{\rm l})(1 - \alpha + \alpha c^{\rm l}) \dfrac{\partial P}{\partial x_i},
\end{equation}
where $\hat{C}^{\rm P,anti}_{\rm seg}$ is a dimensionless quantity, which can depend on additional dimensionless parameters, and the pre-factor $({\bar{d}^2 \dot{\gamma}}/{P})c^{\rm l}(1-c^{\rm l})(1 - \alpha + \alpha c^{\rm l})$ has the same form as that used in the in-plane model summarized in Section~\ref{sec:inplane}, wherein $\alpha$ is a constant parameter. 

We hypothesize that the quantity $\hat{C}^{\rm P,anti}_{\rm seg}$ depends on the inertial number $I$ (as does $\hat{C}^{\rm anti}_{\rm diff}$) and an additional dimensionless parameter $J= {P}/({\bar{d} |\partial P / \partial {\bf x}|})$, which quantifies the ratio of the pressure to the change in pressure over the mean grain size at a point. Inclusion of $J$-dependence in the constitutive equation for the pressure-gradient-driven segregation flux \eqref{eq:seg_flux} allows for potential nonlinear dependence of the flux on the magnitude of the pressure gradient. The procedure to estimate the dimensionless parameter $\hat{C}^{\rm P,anti}_{\rm seg}$ and its functional dependence is as follows. First, we run a given DEM simulation of anti-plane segregation in simple shear flow to steady state, so that the bidisperse mixture is fully segregated, and the total flux at each $z$-position is zero, i.e., $ w^{\rm l}_z = w^{\rm diff}_z+ w^{\rm P}_z =0$. Therefore, the flux balance 
\begin{equation}\label{eq:flux_balance}
-\hat{C}_{\rm diff}^{\rm anti}(I)  \dfrac{\partial c^{\rm l}}{\partial z} = \hat{C}^{\rm P,anti}_{\rm seg}(I,J) \frac{1}{P} c^{\rm l}(1-c^{\rm l})(1 - \alpha + \alpha c^{\rm l}) \dfrac{\partial P}{\partial z}
\end{equation}
is valid at every $z$-position. We find that the steady-state concentration field, and hence the flux balance \eqref{eq:flux_balance}, is independent of the dimensionless strain rate $\dot\gamma\sqrt{\bar{d}_0/G}$ when the other parameters $\{H/\bar{d}_0, c^{\rm l}_0, d^{\rm l}/d^{\rm s}\}$ are fixed. Therefore, \eqref{eq:flux_balance} must be independent of the inertial number, and as a result, the function $\hat{C}^{\rm P,anti}_{\rm seg}(I,J)$ should depend on the inertial number in the same manner as $\hat{C}_{\rm diff}^{\rm anti}(I)$. Accordingly, we take $\hat{C}^{\rm P,anti}_{\rm seg}(I,J)$ to be given by 
\begin{equation}\label{eq:seg_flux2}
\hat{C}^{\rm P,anti}_{\rm seg}(I,J) = \tilde{C}(J)\left[ 1 -  \left(\frac{C_{\infty} - C_0}{C_{\infty} + C_0} \right) \dfrac{(I_{\rm tr}/I)^m - 1}{(I_{\rm tr}/I)^m + 1}\right],
\end{equation}
where $\tilde{C}(J)$ is a function of only $J$. The $I$-dependent function in brackets in \eqref{eq:seg_flux2} is the same as \eqref{eq:cd_anti} but normalized by $(C_\infty + C_0)/2$, and the parameters $\{C_0, C_{\infty}, I_{\rm tr}, m\}$ are the same as those estimated in Section~\ref{sec:diffusion}. 

\begin{figure}[!t]
\centering
\includegraphics{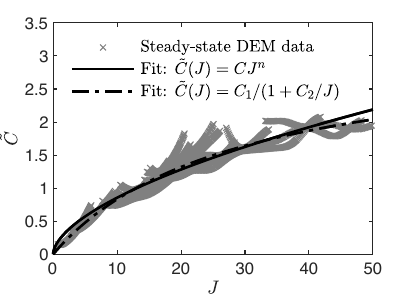} 
\caption{Steady-state collapse of the parameter $\tilde{C}$ \eqref{eq:Ctilde} versus the non-dimensional parameter $J={P}/({\bar{d}|\partial P/\partial z|})$ over $13$ different DEM simulations of anti-plane segregation in simple shear flow. Symbols represent coarse-grained, steady-state DEM field data. The solid black curve is the fit function $\tilde{C} = CJ^n$ with $\{C=0.22, n=0.58\}$, and the dash-dotted black curve is the fit function $\tilde{C} = C_1/(1+C_2/J)$ with $\{C_1 = 3.2, C_2 = 28.62\}$.}\label{fig:Jcollapse}
\end{figure}

Then, plugging \eqref{eq:seg_flux2} into \eqref{eq:flux_balance} and rearranging, we have that, at steady state, $\tilde{C}(J)$ is given by 
\begin{equation}\label{eq:Ctilde}
\tilde{C}(J) = \frac{-\left(C_0 +C_{\infty}\right) P\, \partial c^{\rm l}/ \partial z}{2c^{\rm l}(1- c^{\rm l})(1-\alpha + \alpha c^{\rm l}) \partial P/ \partial z}.
\end{equation}
The field quantities appearing on the right-hand-side of this expression may be obtained by coarse-graining steady-state DEM data, which may be used to determine the function $\tilde{C}(J)$. For a given case of anti-plane segregation in simple shear flow, we consider $z$-positions within the intermediate transition region of the steady-state concentration field where $c^{\rm l}\in (0.2,0.8)$ and calculate the steady-state concentration and pressure fields and their gradients at a given $z$-position from DEM data as described in Appendix~\ref{app:appendix_A}. From the coarse-grained, steady-state DEM data, $\tilde{C}$, using \eqref{eq:Ctilde}, and $J={P}/({\bar{d}|\partial P/\partial z|})$ are calculated at each $z$-position and plotted against one another in Fig.~\ref{fig:Jcollapse}, where each symbol represents a unique $z$-position. We apply this procedure to steady-state DEM data from 13 different cases of anti-plane segregation in simple shear flow by changing the size of the system $H/\bar{d}_0$, the strain rate $\dot\gamma\sqrt{\bar{d}_0/G}$, and the initial conditions $c^{\rm l}_0$, and the data from all cases are included in Fig.~\ref{fig:Jcollapse}. (The grain-size ratio is maintained at $d^{\rm l}/d^{\rm s}=1.5$ throughout.) We consider values of $\alpha$ within the range $\alpha\in (0,0.8)$ and find that the best collapse is obtained for $\alpha=0.4$, which is the same as the value determined for the in-plane mode of segregation in \citet{singh_liu_arxiv}. The resulting steady-state data collapse for $\alpha=0.4$ across all 13 cases is shown in Fig.~\ref{fig:Jcollapse}. The data can be fitted using a power-law function, i.e., $\tilde{C}(J) = C J^{\rm n}$, so that the anti-plane segregation parameter becomes 
\begin{equation}\label{eq:cp_anti}
\hat{C}_{\rm seg}^{\rm P,anti}(I,J) = C J^{\rm n}\left[ 1 -  \left(\frac{C_{\infty} - C_0}{C_{\infty} + C_0} \right) \dfrac{(I_{\rm tr}/I)^m - 1}{(I_{\rm tr}/I)^m + 1}\right],
\end{equation}
where the associated parameters are estimated to be $\{C=0.22, n=0.58\}$. The fitted power-law function is shown by the solid black curve in Fig.~\ref{fig:Jcollapse}. (An alternative fitting function $\tilde{C}(J) = C_1/(1+C_2/J)$ with constants $\{C_1,C_2\}$ based on the works of \citet{trewhela2021experimental} and \citet{barker2021coupling} is discussed in Appendix~\ref{app:appendix_B} and shown by the dash-dotted black curve in Fig.~\ref{fig:Jcollapse}.) The role of $J$-dependence in the segregation flux \eqref{eq:seg_flux} and its effect on the predicted concentration fields is discussed further in Section~\ref{sec:validation}. 

\section{Validation tests in the transient regime}\label{sec:validation}
In Section \ref{sec:diffusion}, we studied anti-plane diffusion in simple shear flow by considering an initially segregated mixture and introduced four parameters $\{C_0, C_{\infty}, I_{\rm tr}, m\}$ to characterize the rate-dependence of the anti-plane mode of diffusion. Then, in Section \ref{sec:segregation}, we used steady-state DEM data from simple shear flow in the presence of an anti-plane gravitational pressure gradient to inform and assess a constitutive equation for the pressure-gradient-driven segregation flux in the anti-plane mode. We introduced three parameters $\{C, n, \alpha\}$ to characterize the anti-plane mode of the segregation flux. In this section, we compare continuum model predictions of the transient evolution of the large-grain concentration field against transient DEM data for several cases of anti-plane segregation in simple shear flow. Throughout, we continue to use the following fixed set of parameters for frictional spheres:
\begin{equation}\label{eq:parameters}
\{C_0 = 0.104, C_{\infty} = 0.057, I_{\rm tr}=0.00744, m = 3.23, C=0.22, n=0.58, \alpha=0.4\}.
\end{equation}

First, we briefly recap how the continuum model is solved to obtain the transient evolution of the $c^{\rm l}$ field in anti-plane segregation in simple shear flow. The large-grain concentration field $c^{\rm l}$ is governed by the following PDE: 
\begin{equation}\label{eq:seg_evol}
\frac{\partial c^{\rm l}}{\partial t} + \frac{\partial }{\partial z} \left( -\hat{C}_{\rm diff}^{\rm anti} \bar{d}^2 \dot{\gamma}  \frac{\partial c^{\rm l}}{\partial z}  - \hat{C}^{\rm P,anti}_{\rm seg} \frac{\bar{d}^2 \dot{\gamma}}{P} c^{\rm l}(1- c^{\rm l})(1- \alpha + \alpha c^{\rm l}) \frac{\partial P}{\partial z} \right) =0,
\end{equation}
where $\hat{C}_{\rm diff}^{\rm anti}(I)$ is given by \eqref{eq:cd_anti} and $\hat{C}^{\rm P,anti}_{\rm seg}(I,J)$ is the dimensionless function depending on $I$ and $J$ given in \eqref{eq:cp_anti}, and  $\bar{d} = c^{\rm l} d^{\rm l} + (1- c^{\rm l}) d^{\rm s} $. As discussed in Section~\ref{sec:antiplaneseg}, the pressure field is hydrostatic (varying linearly with $z$), so that $\partial P/\partial z$ is constant, and to avoid a singularity at the top free surface, a small constant $(1/4)(\partial P/\partial z)\bar{d}_0$ is added to the pressure field. Therefore, we have that $(1/P)\partial P/\partial z = 1/(z + (1/4)\bar{d}_0)$ in the last term of \eqref{eq:seg_evol}, and moreover, the dimensionless quantity $J$ is $J=(z + (1/4)\bar{d}_0)/\bar{d}$. The pressure appearing in the inertial number $I=\dot\gamma \bar{d}\sqrt{\rho_{\rm s}/P}$ is obtained from the coarse-grained DEM data to control for the normal stress differences, and the strain rate $\dot{\gamma}$ is prescribed and spatially uniform. Regarding the boundary conditions, we impose no-flux boundary conditions at the top free surface as well as on the bottom frictionless surface, i.e., $w^{\rm l}_z = - \hat{C}_{\rm diff}^{\rm anti} \bar{d}^2 \dot{\gamma} ({\partial c^{\rm l}}/{\partial z})  - \hat{C}^{\rm P,anti}_{\rm seg} ({\bar{d}^2 \dot{\gamma}}/{P}) c^{\rm l}(1- c^{\rm l})(1- \alpha + \alpha c^{\rm l}) ({\partial P}/{\partial z}) =0$ at $z=0$ and $H$. For the initial condition $c^{\rm l}_0(z)=c^{\rm l}(z, t=0)$, we use the coarse-grained field from the initial DEM configuration for each case and use it as the initial condition in the corresponding continuum simulation. 

\begin{figure}[!t]
\centering
\includegraphics{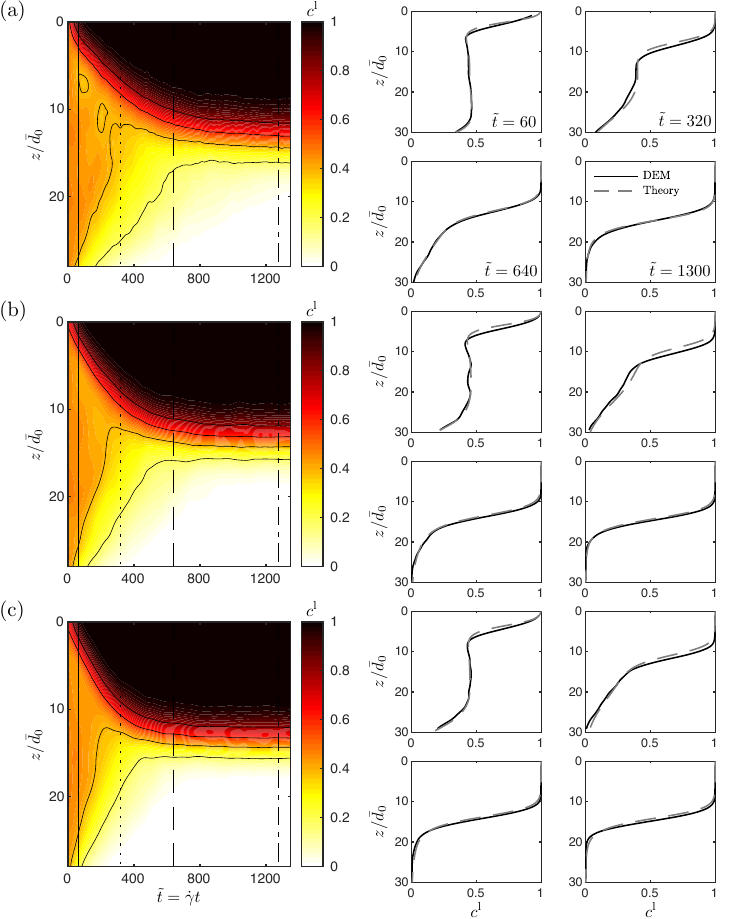} 
\caption{Comparisons of continuum model predictions with corresponding DEM simulation results for the transient evolution of the segregation dynamics for three cases of anti-plane segregation in simple shear flow of bidisperse spheres: (a) Base case $\{H/\bar{d}_0 =30, \dot{\gamma} \sqrt{\bar{d}_0/G} = 0.06, c^{\rm l}_0=0.50,  d^{\rm l}/d^{\rm s} =1.5\}$; (b) Lower strain-rate case I $\{H/\bar{d}_0 =30, \dot{\gamma} \sqrt{\bar{d}_0/G} = 0.03, c^{\rm l}_0=0.50,  d^{\rm l}/d^{\rm s} =1.5\}$; (c) Lower strain-rate case II $\{H/\bar{d}_0 =30, \dot{\gamma} \sqrt{\bar{d}_0/G} = 0.015, c^{\rm l}_0=0.50,  d^{\rm l}/d^{\rm s} =1.5\}$. For each case, spatiotemporal contours of the evolution of $c^{\rm l}$ measured in the DEM simulations are shown on the left. Comparisons of the DEM simulations (solid black lines) and continuum model predictions (dashed gray lines) of the $c^{\rm l}$ field are shown on the right at four time snapshots representing different stages of the segregation process, indicated by the vertical lines on the contour plots for each case.}\label{fig:transient1}
\end{figure}

Next, we compare continuum model predictions of the transient evolution of the segregation process against DEM data for different cases of anti-plane segregation in simple shear flow. In the first set of comparisons, we vary the applied strain rate and consider (1) the base case $\{H/\bar{d}_0 =30$, $\dot{\gamma} \sqrt{\bar{d}_0/G} = 0.06$, $c^{\rm l}_0=0.50$, $d^{\rm l}/d^{\rm s} =1.5\}$, (2) lower strain-rate case I $\{H/\bar{d}_0 =30$, $\dot{\gamma} \sqrt{\bar{d}_0/G} = 0.0$3, $c^{\rm l}_0=0.50$,  $d^{\rm l}/d^{\rm s} =1.5\}$, and (3) lower strain-rate case II $\{H/\bar{d}_0 =30$, $\dot{\gamma} \sqrt{\bar{d}_0/G} = 0.015$, $c^{\rm l}_0=0.50$, $d^{\rm l}/d^{\rm s} =1.5\}$, as shown in Figs.~\ref{fig:transient1}(a), (b) and (c), respectively. Spatiotemporal contours of the evolution of the $c^{\rm l}$ field measured in each DEM simulation are shown on the left, and comparisons between continuum model predictions and DEM simulation data are plotted on the right for four different snapshots in time corresponding to different stages of the segregation process (short time, moderate time, long time, and steady-state), as indicated by the vertical lines on the contour plots for each case. The dimensionless times for each snapshot are given in the bottom right corner of each plot in Fig.~\ref{fig:transient1}(a) and correspond to total shear strains of $\tilde{t} = \dot\gamma t = 60$, 320, 640, and 1300 in all three cases. The steady-state concentration fields at $\tilde{t}=1300$ across all three cases are nearly identical, which is why it was enforced that the steady-state flux balance \eqref{eq:flux_balance} should be independent of the strain rate. While the overall segregation process proceeds faster with increasing strain rate, this dependence is not simply linear. This may be seen by comparing the concentration fields across all three cases at a fixed non-steady-state shear strain. It is clear that the lowest strain-rate case (Fig.~\ref{fig:transient1}(c)) evolves fastest in shear strain to its steady state. This is indicative of the increased rate of diffusion (and hence segregation) at low inertial number (Fig.~\ref{fig:diffusion}(d)). Since this effect has been accounted for in the constitutive equations for the anti-plane modes of diffusion (through \eqref{eq:cd_anti}) and segregation (through \eqref{eq:cp_anti}), the continuum model predicts this characteristic quite well, as can be seen in the transient comparisons. 

\begin{figure}[!t]
\centering
\includegraphics{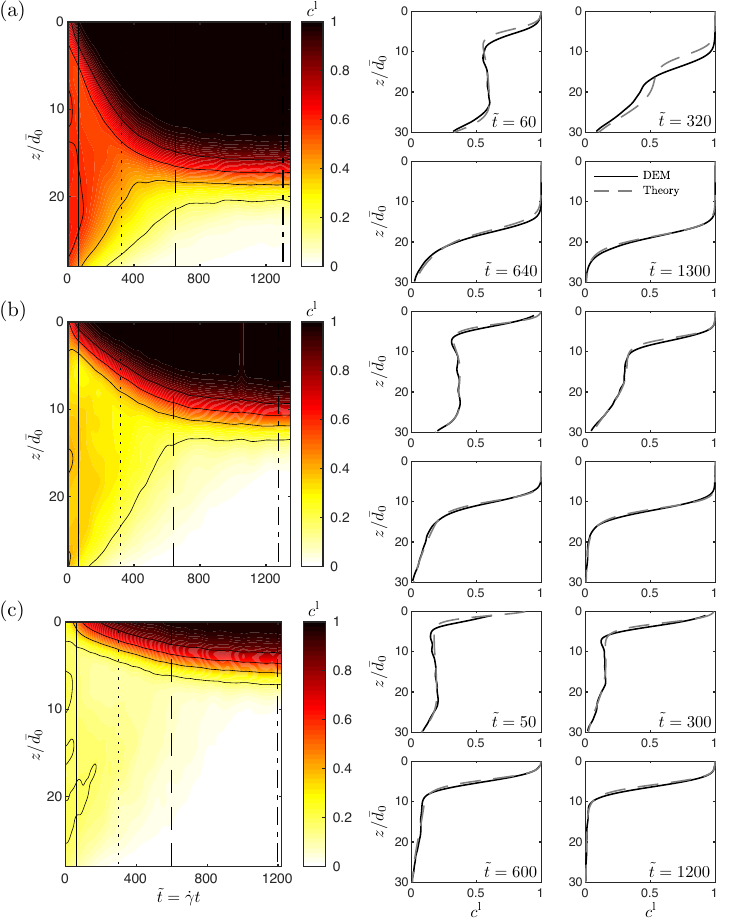} 
\caption{Comparisons of continuum model predictions with corresponding DEM simulation results for the transient evolution of the segregation dynamics for three additional cases of anti-plane segregation in simple shear flow of bidisperse spheres: (a) Higher large-grain concentration case $\{H/\bar{d}_0 =30, \dot{\gamma} \sqrt{\bar{d}_0/G} = 0.06, c^{\rm l}_0=0.60,  d^{\rm l}/d^{\rm s} =1.5\}$; (b) Lower large-grain concentration case I $\{H/\bar{d}_0 =30, \dot{\gamma} \sqrt{\bar{d}_0/G} = 0.06, c^{\rm l}_0=0.40,  d^{\rm l}/d^{\rm s} =1.5\}$; (c) Lower large-grain concentration case II $\{H/\bar{d}_0 =30, \dot{\gamma} \sqrt{\bar{d}_0/G} = 0.06, c^{\rm l}_0=0.20,  d^{\rm l}/d^{\rm s} =1.5\}$. Results are organized as described in the caption of Fig.~\ref{fig:transient1}.}\label{fig:transient2}
\end{figure}

\begin{figure}[!t]
\centering
\includegraphics{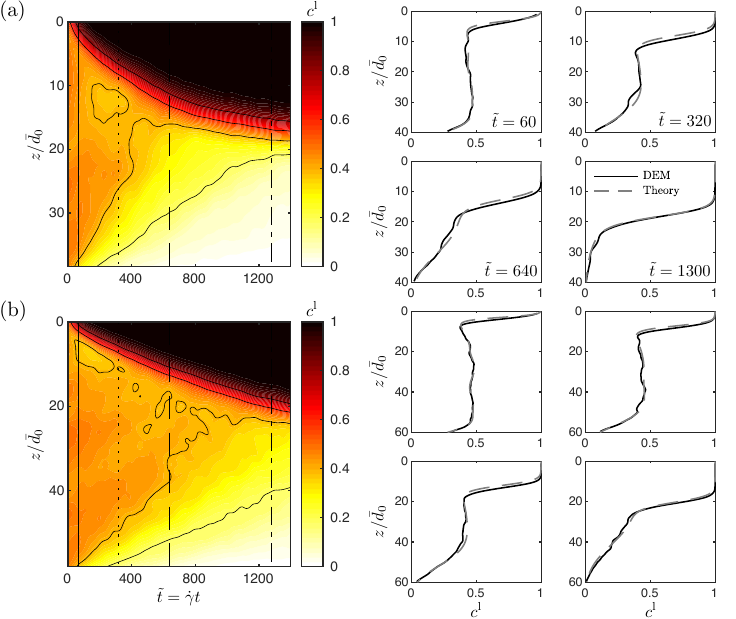} 
\caption{Comparisons of continuum model predictions with corresponding DEM simulation results for the transient evolution of the segregation dynamics for two additional cases of anti-plane segregation in simple shear flow of bidisperse spheres: (a) Deeper layer case I $\{H/\bar{d}_0 =40, \dot{\gamma} \sqrt{\bar{d}_0/G} = 0.06, c^{\rm l}_0=0.50,  d^{\rm l}/d^{\rm s} =1.5\}$; (b) Deeper layer case II $\{H/\bar{d}_0 =60, \dot{\gamma} \sqrt{\bar{d}_0/G} = 0.06, c^{\rm l}_0=0.50,  d^{\rm l}/d^{\rm s} =1.5\}$. Results are organized as described in the caption of Fig.~\ref{fig:transient1}.}\label{fig:transient3}
\end{figure}

Next, we vary the initial concentration of large grains and compare continuum model predictions against DEM simulation data. We consider the following three cases: (1) higher large-grain concentration case $\{H/\bar{d}_0 =30$, $\dot{\gamma} \sqrt{\bar{d}_0/G} = 0.06$, $c^{\rm l}_0=0.60$,  $d^{\rm l}/d^{\rm s} =1.5\}$, (2) lower large-grain concentration case I $\{H/\bar{d}_0 =30$, $\dot{\gamma} \sqrt{\bar{d}_0/G} = 0.06$, $c^{\rm l}_0=0.40$,  $d^{\rm l}/d^{\rm s} =1.5\}$, and (3) lower large-grain concentration case II $\{H/\bar{d}_0 =30$, $\dot{\gamma} \sqrt{\bar{d}_0/G} = 0.06$, $c^{\rm l}_0=0.20$,  $d^{\rm l}/d^{\rm s} =1.5\}$, as shown in Figs.~\ref{fig:transient2}(a), (b), and (c), respectively. Spatiotemporal contours of the evolution of the $c^{\rm l}$ field measured in each DEM simulation are shown on the left, and comparisons of the continuum model predictions against DEM simulation data are plotted on the right for four time instants, as indicated by the vertical lines on the contour plots for each case. Starting with the higher large-grain concentration case, the steady-state transition zone between the large-grain region and the small-grain region is pushed farther beneath the free surface, as can be seen on the bottom right of Fig.~\ref{fig:transient2}(a). As the initial concentration of large grains is reduced, the steady-state transition zone moves closer to the free surface, as seen in the bottom right plots of Figs.~\ref{fig:transient2}(b) and (c). Finally, we vary the depth of the layer and consider two additional cases: (1) deeper layer case I $\{H/\bar{d}_0 =40$, $\dot{\gamma} \sqrt{\bar{d}_0/G} = 0.06$, $c^{\rm l}_0=0.50$,  $d^{\rm l}/d^{\rm s} =1.5\}$ and (2) deeper layer case II $\{H/\bar{d}_0 =60$, $\dot{\gamma} \sqrt{\bar{d}_0/G} = 0.06$, $c^{\rm l}_0=0.50$,  $d^{\rm l}/d^{\rm s} =1.5\}$, as shown in Figs.~\ref{fig:transient3}(a) and (b). Spatiotemporal contours of the evolution of the $c^{\rm l}$ field measured in the DEM simulations are again shown on the left, and profiles of the continuum model predictions compared against DEM simulation data at four time instants are shown on the right. For the deeper layer cases, the steady-state transition zone is farther beneath the free surface compared to the base case of Fig.~\ref{fig:transient1}(a), and it takes more time for the segregation process to reach its steady state. In fact, for deeper layer case II, the steady state is not reached within the simulation time window considered in Fig.~\ref{fig:transient3}(b). In sum, increasing the initial large-grain concentration or the thickness of the layer both result in the steady-state transition zone being farther beneath the free surface, and the continuum model can capture both the transient evolution of the $c^{\rm l}$ field and its steady-state quite well across all of these cases.

A collective view of Figs.~\ref{fig:transient2} and \ref{fig:transient3} reveals that the width of the steady-state transition region increases slightly with its depth beneath the free surface. This is because the pressure-gradient-driven segregation flux attenuates with pressure while the diffusion flux does not, so that the influence of the diffusion flux is comparatively stronger than that of the segregation flux as the depth beneath the free surface increases, leading to a wider transition width. The $J$-dependence of the segregation flux in \eqref{eq:cp_anti} is crucial to capture this effect quantitatively and actually serves to make the segregation flux less dependent on the pressure. Without $J$-dependence, the segregation flux \eqref{eq:segP_eqn} scales as $P^{-1}$ for a fixed magnitude of the pressure gradient, while with the $J$-dependent prefactor \eqref{eq:cp_anti}, the segregation flux scales as $P^{-(1-n)}$, where we have determined that $n=0.58$ for frictional spheres based on the steady-state DEM data of Fig.~\ref{fig:Jcollapse}, so that the segregation flux scales as $P^{-0.42}$. Neglecting $J$-dependence would lead to the segregation flux attenuating too strongly as the pressure increases with depth beneath the free surface, resulting in predicted transition widths that are too sharp near the free surface and too wide deep beneath the free surface. This effect is evident in the transient evolution of the $c^{\rm l}$ field predicted by the $J$-independent, ``in-plane'' version of the continuum model in Fig.~\ref{fig:inplane_model} for the base case, in which the predicted transition width is too sharp at early times when the transition region is near the free surface and too wide at late times when the transition region has moved deeper beneath the free surface. In conclusion, the proposed continuum model, which incorporates $I$- and $J$-dependence in the constitutive equations for the anti-plane fluxes, is capable of quantitatively capturing the segregation dynamics in anti-plane segregation in simple shear flow, including the rate at which the concentration field approaches steady state and the width of the transition region, across various cases subject to changes in strain rate, initial conditions, and layer size. 

\section{Discussion} \label{sec:discussion}

\subsection{Diffusion flux based on mean square displacement data}\label{sec:MSD_anti_plane_diffusion}
In Section~\ref{sec:diffusion}, we characterized the diffusion flux by starting with a fully segregated granular mixture and examining the dynamics of mixing under conditions in which diffusion is dominant. However, it is common in the literature \citep[e.g.,][]{bancroft2021drag} to characterize the diffusion flux by calculating the mean square displacement (MSD) of the granular mixture. In this section, we follow this process to characterize the diffusion flux and assess whether the resulting flux constitutive equations can capture the dynamics of the $c^{\rm l}$ field in the anti-plane mode of diffusion and segregation. 

\begin{figure}[!t]
\centering
\includegraphics{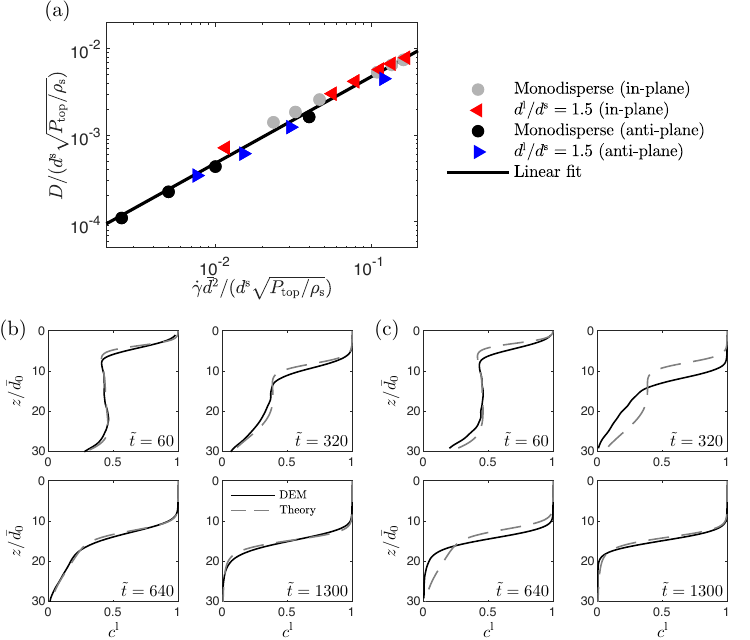} 
\caption{(a) The binary diffusion coefficient $D$, calculated using the mean square displacement, plotted versus $\dot{\gamma} \bar{d}^2$ in homogeneous, steady simple shear DEM simulations. Each symbol represents $D$ calculated from a distinct DEM simulation. Both axes are normalized by $d^{\rm s} \sqrt{P_{\rm top}/\rho_{\rm s}}$. The solid line is the best fit of a linear relation with a slope of $C_{\rm diff} = 0.048$. Comparisons between continuum model predictions (dashed gray lines) and DEM simulation results (solid black lines) of the $c^{\rm l}$ field at four time snapshots using a constitutive equation for the diffusion flux based on mean square displacement data for (b) the base case of anti-plane segregation in simple shear flow $\{H/\bar{d}_0 =30, \dot{\gamma} \sqrt{\bar{d}_0/G} = 0.06, c^{\rm l}_0=0.50,  d^{\rm l}/d^{\rm s} =1.5\}$ and (c) the lower strain-rate case II $\{H/\bar{d}_0 =30, \dot{\gamma} \sqrt{\bar{d}_0/G} = 0.015, c^{\rm l}_0=0.50,  d^{\rm l}/d^{\rm s} =1.5\}$.}\label{fig:msd}
\end{figure}

To this end, we consider both the in-plane and anti-plane modes of diffusion in DEM simulations of simple shear. In our prior work \citep{liu2023coupled}, the in-plane MSD was calculated from DEM data for steady, simple shear flows of the same dense granular system (Appendix~\ref{app:appendix_A}) to determine the binary diffusion coefficient $D$ for both bidisperse mixtures of spheres and for the monodisperse case. We follow an analogous procedure here to calculate the diffusion coefficient in the anti-plane mode. DEM simulations of simple shear flow are performed as described in Section~\ref{sec:diffusion} for a well-mixed bidisperse granular system with $d^{\rm l}/d^{\rm s}=1.5$ as well as for the monodisperse case. The MSD of a system of $N$ particles along the anti-plane direction (i.e., the $z$-direction) is calculated as a function of time \citep[e.g.,][]{natarajan1995,campbell1997self,utter2004,fry2019diffusion,bancroft2021drag} according to
\begin{equation}\label{eq:msd}
\text{MSD}(t)=\dfrac{1}{N} \sum_{n=1}^N(z_n(t)-z_n(0))^2 = 2Dt,
\end{equation}
where $z_n(t)$ is the $z$-coordinate of the $n$th grain at time $t$. The binary diffusion coefficient $D$ may then be inferred from the slope of the MSD in time. To avoid boundary effects in the calculation of the MSD, grains that are initially within $10\bar{d}_0$ of either the top or bottom boundary are excluded from the system of $N$ particles used to calculate the MSD, leaving a set of $N\approx 3000$ grains. The normalized binary diffusion coefficient $D$ is plotted versus $\dot{\gamma} \bar{d}^2$ (with both quantities normalized by $d^{\rm s} \sqrt{P_{\rm top}/\rho_{\rm s}}$) for different strain rates in Fig.~\ref{fig:msd}(a) for both modes of diffusion and both bidisperse and monodisperse granular systems. The DEM data for the in-plane mode of diffusion in Fig.~\ref{fig:msd}(a) is the data reported in Fig.~3(b) of \citet{liu2023coupled}. The DEM data for the binary diffusion coefficient collapses to a nearly linear relation with $D\sim \dot{\gamma}\bar{d}^2$ across the range of strain rates considered and for both monodisperse and bidisperse systems in in-plane as well as anti-plane modes of diffusion. The solid black line in Fig.~\ref{fig:msd}(a) is a linear fit that estimates the diffusion parameter as $C_{\rm diff} = 0.048$, which is very close to the estimated value of $C_{\rm diff}^{\rm in} = 0.045$ determined in \citet{liu2023coupled}.

The diffusion parameter $C_{\rm diff}$ estimated via the MSD approach in Fig.~\ref{fig:msd}(a) is approximately rate-independent, i.e, it is only weakly dependent upon the strain rate (or inertial number), which contradicts the previously observed $I$-dependence at the continuum scale discussed in Section~\ref{sec:diffusion} and illustrated in Fig.~\ref{fig:diffusion}(d). Therefore, we test this constant value of $C_{\rm diff}=0.048$ in the segregation dynamics equation \eqref{eq:seg_evol}. In doing so, $I$-dependence is neglected in both the diffusion and segregation parameters, so that the segregation dynamics equation becomes 
\begin{equation}\label{eq:seg_evol2}
\frac{\partial c^{\rm l}}{\partial t} + \frac{\partial }{\partial z} \left( -{C}_{\rm diff} \bar{d}^2 \dot{\gamma}  \frac{\partial c^{\rm l}}{\partial z}  - C J^n \frac{\bar{d}^2 \dot{\gamma}}{P} c^{\rm l}(1- c^{\rm l})(1- \alpha + \alpha c^{\rm l}) \frac{\partial P}{\partial z} \right) =0,
\end{equation}
with the reduced parameter set $\{C_{\rm diff}=0.048, C=0.22, n=0.58, \alpha=0.4\}$. This approach maintains the steady-state collapse of Fig.~\ref{fig:Jcollapse}. We then numerically calculate the evolution of the large-grain concentration field using \eqref{eq:seg_evol2} accompanied by the same boundary and initial conditions as in the preceding sections. First, we compare predictions of the continuum model with DEM simulation data for the base case of anti-plane segregation in simple shear flow (Fig.~\ref{fig:transient1}(a)) as shown in Fig.~\ref{fig:msd}(b). The continuum model predictions using the constant diffusion parameter capture the DEM simulation results reasonably well for this case. However, for the lower strain-rate case II (Fig.~\ref{fig:transient1}(c)), as shown in Fig. \ref{fig:msd}(c), continuum model predictions using the constant diffusion parameter lag behind the DEM data in time. This indicates that the diffusion parameter is indeed rate-dependent, and at low strain rates, its value is higher than what is inferred from  MSD data. This is consistent with the estimates of the diffusion parameter shown in Fig. \ref{fig:diffusion}(d). We leave this discrepancy between the MSD-based estimate and the continuum-model-fitting-based estimate of the diffusion parameter as an open problem for future work. However, we have established in this study that the anti-plane diffusion parameter is rate-dependent in contrast to the in-plane diffusion parameter, which is nearly rate-independent. This implies that the diffusion process in general is mode-dependent and to account for the mode-dependence, we propose a strategy for generalizing the diffusion parameter in the next section. 

\subsection{Generalized three-dimensional constitutive relations}\label{sec:3D_constitutive_form}
In Section~\ref{sec:antiplanefluxes}, we proposed constitutive equations for the anti-plane modes of the diffusion flux and the pressure-gradient-driven segregation flux. Both the anti-plane diffusion and segregation fluxes take different forms from the in-plane modes studied in our prior works \citep{liu2023coupled,singh_liu_arxiv} and described in Section~\ref{sec:inplane}. In this section, we propose a strategy for generalizing the constitutive forms for both fluxes, which takes into account both the in-plane and the anti-plane modes of diffusion and pressure-gradient-driven segregation. 

In order to propose a generalized constitutive form for the diffusion flux, we denote the unit vector along the direction of the concentration gradient as $n^{\rm c}_i = ({\partial c^{\rm l}/\partial x_i})/{|\partial c^{\rm l}/\partial {\bf x}|}$ and the flow direction tensor as $N_{ij} = {D_{ij}}/{|{\bf D}|}$. Next, we introduce a combined scalar invariant $K^{\rm c} = n^{\rm c}_i N_{ij} N_{jk} n^{\rm c}_k$, which is used to quantify whether diffusion is in its in-plane or anti-plane mode. Straightforward calculations described in Appendix~\ref{app:appendix_C} show that for the in-plane mode of diffusion, the scalar invariant is $K^{\rm c} = 1/2$, and for the anti-plane mode of diffusion, the scalar invariant is $K^{\rm c} = 0$. This feature indicates that $K^{\rm c}$ is a suitable invariant for capturing mode-dependence, and we define a generalized diffusion parameter through a linear combination of $\hat{C}_{\rm diff}^{\rm anti}$ and $C_{\rm diff}^{\rm in}$ based on the invariant $K^{\rm c}$ as follows:
\begin{equation}\label{eq:gen_diff_coeff}
\hat{C}_{\rm diff}(I,K^{\rm c}) = \hat{C}^{\rm anti}_{\rm diff}(I) + 2 K^{\rm c} \left( C^{\rm in}_{\rm diff} - \hat{C}^{\rm anti}_{\rm diff}(I)\right),
\end{equation}
where $C^{\rm in}_{\rm diff}$ is the constant diffusion parameter for the in-plane mode of diffusion determined in \citet{liu2023coupled} for frictional spheres. In \eqref{eq:gen_diff_coeff}, $\hat{C}_{\rm diff} = C^{\rm in}_{\rm diff}$ for the in-plane mode when $K^{\rm c}=1/2$, and  $\hat{C}_{\rm diff} = \hat{C}^{\rm anti}_{\rm diff}$ for the anti-plane mode when $K^{\rm c} = 0$. The generalized diffusion flux may then be written as $w^{\rm diff}_i = -\hat{C}_{\rm diff}(I,K^{\rm c})  \bar{d}^2 \dot{\gamma} ({\partial c^{\rm l}}/{\partial x_i})$. 

Similarly, in order to propose a generalized constitutive equation for the pressure-gradient-driven segregation flux, we denote the unit vector along the direction of the pressure gradient as $n^{\rm P}_i = ({\partial P/\partial x_i})/{|\partial P/\partial {\bf x}|}$, and introduce another combined scalar invariant $K^{\rm P} = n^{\rm P}_i N_{ij} N_{jk} n^{\rm P}_k$. The generalized pressure-gradient segregation parameter is then defined through a linear combination of $\hat{C}^{\rm P,anti}_{\rm seg}$ and $C^{\rm P,in}_{\rm seg}$ as 
\begin{equation}\label{eq:gen_seg_coeff}
\hat{C}^{\rm P}_{\rm seg}(I, J, K^{\rm P}) = \hat{C}_{\rm seg}^{\rm P,anti}(I, J) + 2 K^{\rm P} \left( C_{\rm seg}^{\rm P,in} - \hat{C}_{\rm seg}^{\rm P,anti}(I, J)\right),
\end{equation}
where $C_{\rm seg}^{\rm P,in}$ is the constant in-plane segregation parameter, so that the generalized pressure-gradient-driven segregation flux is 
\begin{equation}\label{eq:gen_seg_flux}
w^{\rm P}_i = -\hat{C}^{\rm P}_{\rm seg}(I,J,K^{\rm P})  \frac{ \bar{d}^2 \dot\gamma}{P} c^{\rm l} (1- c^{\rm l})(1-\alpha +\alpha c^{\rm l})\frac{\partial P}{\partial { x}_i}.
\end{equation}
Analogous to the discussion of the preceding paragraph, for the in-plane mode of segregation, $K^{\rm P}=1/2$ and $\hat{C}^{\rm P}_{\rm seg} = C_{\rm seg}^{\rm P,in}$, and for the anti-plane mode of segregation, $K^{\rm P}=0$ and $\hat{C}^{\rm P}_{\rm seg} = \hat{C}_{\rm seg}^{\rm P,anti}$. 

Putting the generalized flux constitutive equations together, the generalized form of the segregation dynamics equation can be written as
\begin{equation}\label{eq:seg_evol_3D}
\frac{D c^{\rm l}}{D t} + \frac{\partial }{\partial x_i} \left( -\hat{C}_{\rm diff}(I,K^{\rm c}) \bar{d}^2 \dot{\gamma} \frac{\partial c^{\rm l}}{\partial x_i}  - \hat{C}^{\rm P}_{\rm seg}(I,J,K^{\rm P}) \frac{\bar{d}^2 \dot{\gamma}}{P} c^{\rm l}(1- c^{\rm l})(1- \alpha + \alpha c^{\rm l}) \frac{\partial P}{\partial x_i} \right) =0,
\end{equation}
where $\hat{C}_{\rm diff}$ and $\hat{C}^{\rm P}_{\rm seg}$ are given by \eqref{eq:gen_diff_coeff} and \eqref{eq:gen_seg_coeff}, respectively. The diffusion parameter $\hat{C}_{\rm diff}$ is described by the parameter set $\{ C_{\rm diff}^{\rm in}, C_0, C_{\infty}, I_{\rm tr}, m\}$, and the segregation parameter $\hat{C}^{\rm P}_{\rm seg}$ is described by the parameter set $\{ C^{\rm P,in}_{\rm seg}, C, n\}$. In the limits of pure in-plane and pure anti-plane modes, the generalized segregation dynamics equation \eqref{eq:seg_evol_3D} will reduce to the appropriate forms that have been tested in their respective modes, and future work will test the generalized model \eqref{eq:seg_evol_3D} in more complex flow configurations that involve mixed-mode diffusion and segregation, such as split-bottom flow \citep{hill08,fan10}.

\section{Concluding remarks}\label{sec:conclusion}
In this paper, we have studied diffusion and size segregation in their anti-plane modes in dense, bidisperse granular mixtures of spheres. In DEM simulations, we observed that the diffusion and pressure-gradient-driven segregation fluxes are mode-dependent in nature, and the previously-developed, in-plane size-segregation model of \citet{singh_liu_arxiv} is unable to capture the segregation dynamics in the anti-plane mode of segregation. Guided by DEM simulations, we proposed phenomenological constitutive equations for the anti-plane diffusion and segregation fluxes. We found that the anti-plane diffusion parameter $\hat{C}_{\rm diff}^{\rm anti}$ is rate dependent and depends on the inertial number $I$, and we proposed a phenomenological fitting function \eqref{eq:cd_anti} to capture the $I$-dependence. Moreover, we found that the pressure-gradient-driven segregation flux depends on the magnitude of the pressure gradient in a nonlinear fashion, and we proposed a constitutive equation for the anti-plane segregation parameter $\hat{C}^{\rm P,anti}_{\rm seg}$ \eqref{eq:cp_anti} that captures this dependence using an additional dimensionless parameter $J={P}/{(\bar{d}|\partial P/\partial \textbf{x}|)}$. The newly-developed phenomenological constitutive equations predict the transient evolution of the segregation dynamics quite well across different cases of anti-plane segregation in simple shear flow as the strain rate, initial conditions, and layer size are varied. Finally, we have proposed a strategy for generalizing the constitutive equations for the diffusion and pressure-gradient-driven segregation fluxes that synthesizes the anti-plane flux constitutive equations proposed in this work with the in-plane flux constitutive equations of \citet{singh_liu_arxiv}, which is suitable for more complex, three-dimensional flows. 

Although the proposed continuum model performs well in predicting the segregation dynamics in anti-plane segregation in simple shear flow, several important questions remain. First, the present work has focused on a single grain-size ratio of $d^{\rm l}/d^{\rm s}=1.5$. It has been established that the in-plane pressure-gradient-driven segregation flux should depend upon the grain-size ratio \citep[e.g.,][]{tunuguntla2014mixture,schlick2015,jones2018asymmetric,trewhela2021experimental}, and it remains to probe this dependence in the anti-plane mode. Second, future work is required to investigate the discrepancy between the MSD-based estimate and the continuum-model-fitting based estimate of the diffusion parameter in the anti-plane mode of diffusion, as discussed in Section~\ref{sec:MSD_anti_plane_diffusion}. Finally, the generalized constitutive equations that account for both in-plane and anti-plane modes of diffusion and segregation still need to be tested in more complex flow configurations---for example,  annular shear flow with gravity, split bottom flow \citep{hill08,fan10}, or flow in a blade mixer \citep{yang2022continuum}. To achieve this, it is necessary to integrate the proposed generalized segregation model with rheological constitutive equations (e.g., the nonlocal granular fluidity model \citep{kamrin2012,henann2013}) to develop a coupled continuum model that is able to predict the flow and segregation dynamics simultaneously and to develop a robust numerical formulation to solve the coupled system of continuum equations in complex geometries. These points will be pursued in future work.

\section*{Acknowledgements}
This work was supported by funds from NSF-CBET-1552556.

\appendix
\section{Simulated granular system and averaging methods} \label{app:appendix_A}
We consider granular systems consisting of dense, bidisperse mixtures of spheres. The mean large-grain diameter is $d^{\rm l}=3\, {\rm mm}$, and the mean small-grain diameter is $d^{\rm s} = 2 \, {\rm mm}$, so that the grain-size ratio $d^{\rm l}/d^{\rm s}=1.5$ is held constant throughout. We impose a polydispersity of $\pm 10 \%$ to the mean diameter of each species to prevent crystallization. The grain-material density is $\rho_{\rm s} = 2450\, {\rm kg/m}^3$. The grain-grain interaction force is given through a contact law that accounts for Hookean elasticity, damping, and sliding friction \citep{silbert2001granular}. The relevant interaction properties are the normal contact stiffness $k_{\rm n}$, the tangential contact stiffness $k_{\rm t}$, the coefficient of restitution $e$, and the inter-particle friction coefficient $\mu_{\rm surf}$. We restrict our attention to the nearly-rigid particle regime and set the normal contact stiffness to be sufficiently high compared to the confining pressure throughout, i.e.,  $k_{\rm n}/P \bar{d}_0 > 10^4$. The other parameters are set as $k_{\rm t}/k_{\rm n}=1/2$, $e=0.1$, and $\mu_{\rm surf} = 0.4$ throughout. Numerical integration of the equations of motion for each grain is performed using the open-source software LAMMPS \citep{lammps}, and the time step of numerical integration is specified to sufficiently small to ensure accuracy and stability of simulation results. 

To extract coarse-grained concentration and pressure fields from DEM simulations, we use a bin-based approach, as described in our prior work \citep{liu2023coupled}. Briefly, this approach utilizes rectangular-cubiodal bins of width $\Delta$ that are centered at a $z$-position and span the simulation domain along the $x$- and $y$-directions. For a snapshot at time $t$, each grain $i$ intersected by the bin is given a weight $V_i$, which is equal to the volume of grain $i$ inside the bin. The sets of large and small grains intersected by the bin are denoted as ${\cal F}^{\rm l}$ and ${\cal F}^{\rm s}$, respectively \citep{tunuguntla2016}. The instantaneous solid volume fractions for large and small grains are $\phi^{\rm l}(z,t) = (\sum_{i \in {\cal F}^{\rm l}} V_i)/V$ and $\phi^{\rm s}(z,t) = (\sum_{i \in {\cal F}^{\rm s}} V_i)/V$, respectively, where $V$ is the volume of the bin. The concentration field for the large grains is calculated as $c^{\rm l}(z,t) = \phi^{\rm l}(z,t)/(\phi^{\rm l}(z,t)+\phi^{\rm s}(z,t))$. The instantaneous stress tensor associated with grain $i$ is $\boldsymbol{\sigma}_i(t) = (\sum_{j\ne i} {\bf r}_{ij}\otimes {\bf f}_{ij})/(\pi d_i^3/6)$, where ${\bf r}_{ij}$ is the position vector from the center of grain $i$ to the center of grain $j$, ${\bf f}_{ij}$ is the contact force applied on grain $i$ by grain $j$, and $d_i$ is the diameter of grain $i$. The instantaneous stress field is $\boldsymbol{\sigma}(z,t) = (\sum_{i \in {\cal F}} V_i\boldsymbol{\sigma}_i(t))/V$, where ${\cal F} = {\cal F}^{\rm l} \cup {\cal F}^{\rm s}$ is the set of all grains intersected by the bin, and the instantaneous pressure field is $P(z,t) = -(1/3){\rm tr}(\boldsymbol{\sigma}(z,t))$. Throughout, we take a bin width of $\Delta = 4 \bar{d}_0$ and a spatial resolution of about $0.2 \bar{d}_0$. For the steady-state collapse of Fig.~\ref{fig:Jcollapse}, we consider DEM data from the spatial region where $c^{\rm l} \in (0.2, 0.8)$. When calculating the quantities appearing in the collapse of Fig.~\ref{fig:Jcollapse}, the instantaneous concentration and pressure fields are spatially smoothed and differentiated using Lucy's quartic kernel function \citep{tunuguntla2016} with a kernel width of $4\bar{d}_0$. To obtain the relevant steady-state field quantities in \eqref{eq:Ctilde}, we consider a time window within the steady-state regime and generate $N=1000$ snapshots of the instantaneous, smoothed fields and arithmetically average these fields over all snapshots to obtain fields that only depend on the spatial coordinate.

\section{Alternative constitutive equation for the segregation flux}\label{app:appendix_B}
The works of \citet{trewhela2021experimental} and \citet{barker2021coupling} suggest a constitutive form for the pressure-gradient-driven segregation flux that depends on $J$ in an alternative manner, adapted to the present work as follows:
\begin{equation}\label{eq:segflux_alt}
w^{\rm P}_i = \dfrac{C_1}{1 + C_2/J} \left[ 1 -  \left(\frac{C_{\infty} - C_0}{C_{\infty} + C_0} \right) \dfrac{(I_{\rm tr}/I)^m - 1}{(I_{\rm tr}/I)^m + 1}\right] \dfrac{\bar{d}^2 \dot{\gamma}}{P} c^{\rm l}(1-c^{\rm l})(1 - \alpha + \alpha c^{\rm l}) \dfrac{\partial P}{\partial x_i},
\end{equation}
where the flux depends on $J$ through the factor $\tilde{C}(J) = C_1/(1+C_2/J)$ rather than $\tilde{C}(J) = C J^n$, and $\{C_1, C_2\}$ are constant fitting parameters. The prefactor $(C_1/(1+C_2/J))(\bar{d}^2\dot\gamma/P)$ in \eqref{eq:segflux_alt} may be rewritten as $C_1\bar{d}^2\dot\gamma/(P + C_2\bar{d} |\partial P/\partial{\bf x}|)$, and the term in the denominator involving $C_2\bar{d} | \partial P/\partial{\bf x}|$ was included by \citet{trewhela2021experimental} and used by \citet{barker2021coupling} to prevent a singularity when the pressure equals zero, such as at a free surface. (The flux constitutive equation using the factor $CJ^n$ does exhibit a singularity when the pressure equals zero, and in Section~\ref{sec:validation}, we circumvented this issue by adding a small top pressure equal to the weight of a layer of $(1/4)\bar{d}_0$ thickness, i.e., $(1/4)(\partial P/\partial z)\bar{d}_0$, to the pressure field.) Here, we consider the alternative segregation flux constitutive equation \eqref{eq:segflux_alt} and assess its ability to capture segregation dynamics in the anti-plane mode of segregation. 

\begin{figure}[!t]
\centering
\includegraphics{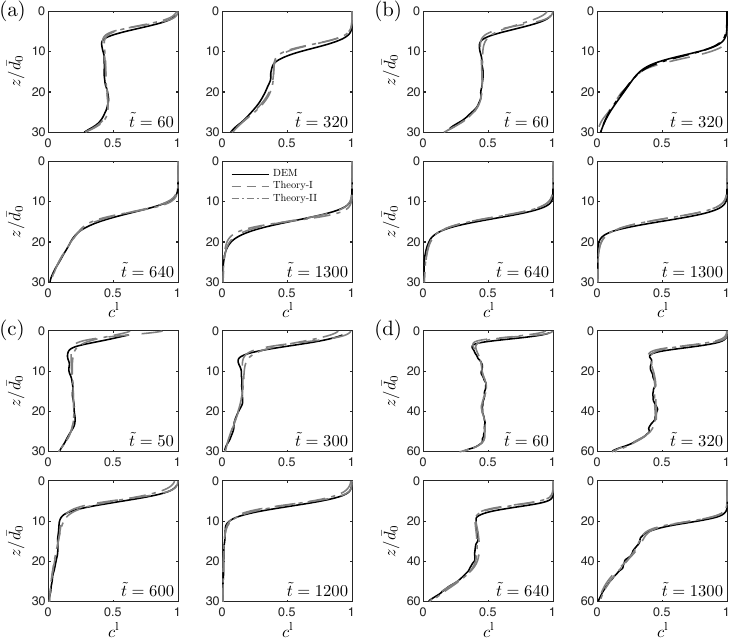} 
\caption{Comparisons of continuum model predictions with corresponding DEM simulation results for the transient evolution of the segregation dynamics for four different cases of anti-plane segregation in simple shear flow of bidisperse spheres: (a) Base case $\{H/\bar{d}_0 =30, \dot{\gamma} \sqrt{\bar{d}_0/g} = 0.06, c^{\rm l}_0=0.50,  d^{\rm l}/d^{\rm s} =1.5\}$; (b) Lower strain-rate case II $\{H/\bar{d}_0 =30, \dot{\gamma} \sqrt{\bar{d}_0/g} = 0.015, c^{\rm l}_0=0.50,  d^{\rm l}/d^{\rm s} =1.5\}$; (c) Lower large-grain concentration case II $\{H/\bar{d}_0 =30, \dot{\gamma} \sqrt{\bar{d}_0/g} = 0.06, c^{\rm l}_0=0.20,  d^{\rm l}/d^{\rm s} =1.5\}$; and (d) Deeper layer case II $\{H/\bar{d}_0 =60, \dot{\gamma} \sqrt{\bar{d}_0/G} = 0.06, c^{\rm l}_0=0.50,  d^{\rm l}/d^{\rm s} =1.5\}$. DEM simulation results are shown using solid black lines. Theory-I, shown using dashed gray lines, corresponds to predictions of the continuum model using the prior constitutive form for the anti-plane segregation flux \eqref{eq:cp_anti}. Theory-II, shown using dash-dotted gray lines, corresponds to predictions of the continuum model using the alternative constitutive form \eqref{eq:segflux_alt}.}\label{fig:altmod}
\end{figure}

With $\tilde{C}(J) = C_1 / (1 + C_2/J)$, the parameters $\{C_1, C_2\}$ may be determined by fitting to the steady-state DEM data from 13 different cases of anti-plane segregation in simple shear flow, and the best fit is shown by the dash-dotted black line in Fig.~\ref{fig:Jcollapse}. The estimated parameters are $\{C_1 = 3.2, C_2 = 28.62\}$. We note that the value of $C_2=28.62$ is quite large compared to the value of 0.2712 used by \citet{trewhela2021experimental} for the purpose of preventing a singularity, so in the anti-plane context, this term plays a key role in capturing the dynamics of segregation and is not simply included to regularize the free-surface singularity. Then, we plug the alternative constitutive form for the pressure-gradient-driven segregation flux in the segregation dynamics equation \eqref{eq:seg_evol} and compare numerical predictions of the dynamics of the $c^{\rm l}$ field against the DEM simulation results. We consider four different cases: (1) the base case, (2) lower strain-rate case II, (3) lower large-grain concentration case II, and (4) deeper layer case II, as shown in Figs.~\ref{fig:altmod}(a), (b), (c), and (d), respectively. The predictions of the alternative constitutive form are shown using dash-dotted gray lines and are as capable of capturing the segregation dynamics as predictions of the prior constitutive form, which are shown using dashed gray lines. 

\section{Generalized diffusion and segregation parameters} \label{app:appendix_C}
In this appendix, we give additional context to the combined invariants used in Section~\ref{sec:3D_constitutive_form} to obtain the generalized diffusion parameter  \eqref{eq:gen_diff_coeff} and the generalized segregation parameter \eqref{eq:gen_seg_coeff}. Here, we focus the discussion on the constitutive equation for the generalized diffusion parameter, but an analogous process may be applied to the generalized segregation parameter. Between our prior work \citep{liu2023coupled} and Section~\ref{sec:diffusion} of the present work, we have observed that the diffusion process depends on the mode of diffusion relative to the plane of shearing (in-plane or anti-plane) and the inertial number $I$. To account for this dependence in a constitutive equation, we consider the flow direction tensor ${\bf N} = {\bf D}/|{\bf D}|$, which is deviatoric ${\rm tr}({\bf N}) = 0$ and of unit magnitude $|{\bf N}| =1$, and the unit vector along the concentration gradient ${\bf n}^{\rm c} = (\partial c^{\rm l}/\partial{\bf x})/|\partial c^{\rm l}/\partial{\bf x}|$. Assuming that the diffusion flux and the concentration gradient act in opposite directions along the same line of action, a generalized constitutive equation for the diffusion flux may then be written in terms of ${\bf N}$ and ${\bf n}^{\rm c}$ as follows: 
\begin{equation}\label{eq:generalized_diff_flux_app}
w^{\rm diff}_i = - \check{C}_{\rm diff} \left(I,{\bf N}, {\bf n}^{\rm c}\right)  \bar{d}^2 \dot{\gamma} \dfrac{\partial c^{\rm l}}{\partial x_i}, 
\end{equation} 
where $\check{C}_{\rm diff} \left(I, {\bf N}, {\bf n}^{\rm c}\right)$ is a scalar function that accounts for the mode-dependent nature of diffusion. The function $\check{C}_{\rm diff} \left(I, {\bf N}, {\bf n}^{\rm c}\right)$ must be an isotropic function of its arguments, so it may be written in terms of the combined invariants associated with the unit flow direction tensor ${\bf N}$ and the unit vector ${\bf n}^{\rm c}$ as follows:
\begin{equation}\label{eq:invariants}
\begin{split}
& K_1 = {\rm tr} (\textbf{N}),\,\,\, K_2 = {\rm tr} (\textbf{N}^2) ,\,\,\, K_3 = {\rm tr} (\textbf{N}^3), \\
& K_4 = \textbf{n}^{\rm c} \cdot \textbf{n}^{\rm c},\,\,\, K_5 = \textbf{n}^{\rm c}\cdot \textbf{N}  \textbf{n}^{\rm c}, \,\,\,  K_6 = \textbf{n}^{\rm c}\cdot  \textbf{N}^2  \textbf{n}^{\rm c},
\end{split}
\end{equation} 
where the invariants $K_1=0,\, K_2=1$ and $K_4=1$ are constants due to the properties of ${\bf N}$ and ${\bf n}^{\rm c}$, and the remaining invariant set reduces to $\{K_3,K_5,K_6\}$. (The invariants are denoted as $\{K_1,K_2,K_3,K_4,K_5,K_6\}$ to avoid confusion with the inertial number $I$ and the dimensionless quantity $J$ introduced in Section~\ref{sec:segregation}.) An isotropic representation of $\check{C}_{\rm diff}\left(I, {\bf N}, {\bf n}^{\rm c}\right)$ may then be written as $\hat{C}_{\rm diff}\left(I,K_3,K_5,K_6\right)$, and the flux constitutive equation \eqref{eq:generalized_diff_flux_app} reduces to 
\begin{equation}\label{eq:rep_thm}
w^{\rm diff}_i = - \hat{C}_{\rm diff} \left(I,K_3,K_5,K_6\right)  \bar{d}^2 \dot{\gamma} \dfrac{\partial c^{\rm l}}{\partial x_i}.
\end{equation}

To further reduce \eqref{eq:rep_thm}, we consider a situation, in which the concentration field only varies along the $z$-direction, so that $[{\bf n}^{\rm c}] = \pm[0\,\,\,\,0\,\,\,\,1]^\top$, and the velocity field corresponds to a steady, viscometric flow of the following form: $v_x(y,z)$, $v_y=0$, and $v_z=0$. The velocity gradient tensor ${\bf L} = {\rm grad} \, {\bf v}$ and the strain-rate tensor ${\bf D} = (1/2) ({\bf L} + {\bf L}^\top) $ are then
\begin{equation}
[\textbf{L}] = \begin{bmatrix}
0 & \partial v_x /\partial y & \partial v_x /\partial z\\
0 & 0 & 0 \\
0 & 0 & 0
\end{bmatrix} \quad \text{and} \quad
[\textbf{D}] = \frac{1}{2}\begin{bmatrix}
0 & \partial v_x /\partial y & \partial v_x /\partial z\\
\partial v_x /\partial y & 0 & 0 \\
\partial v_x /\partial z & 0 & 0
\end{bmatrix}.
\end{equation}
This steady, viscometric flow field involves shear within two planes: the $x$-$y$-plane and the $x$-$z$-plane. The concentration gradient lies within the $x$-$z$-plane, so we denote $\partial v_x/\partial z=\dot\gamma_{\rm in}$. Similarly, the concentration gradient is perpendicular to the $x$-$y$-plane, so we denote $\partial v_x/\partial y=\dot\gamma_{\rm anti}$. The equivalent shear strain rate is $\dot{\gamma} = \sqrt{2} |{\bf D}| = \sqrt{(\partial v_x/ \partial y)^2 + (\partial v_x/ \partial z)^2} =\sqrt{\dot\gamma_{\rm anti}^2 + \dot\gamma_{\rm in}^2}$. The unit flow direction tensor ${\bf N} = {\bf D}/|{\bf D}|$ may then be written as 
\begin{equation}
[{\bf N}] = \frac{\sqrt{2} [{\bf D}]}{\dot{\gamma}} = \frac{1}{\sqrt{2}} \begin{bmatrix}
0 & \dot{\gamma}_{\rm anti} / \dot{\gamma}  & \dot{\gamma}_{\rm in} / \dot{\gamma} \\
\dot{\gamma}_{\rm anti} / \dot{\gamma} & 0 & 0 \\
\dot{\gamma}_{\rm in} / \dot{\gamma} & 0 & 0
\end{bmatrix}, 
\end{equation}
and the tensor ${\bf N}^2$ becomes
\begin{equation}
[{\bf N}^2] = \frac{1}{2}\begin{bmatrix}
\frac{\dot{\gamma}_{\rm anti}^2 + \dot{\gamma}_{\rm in}^2}{\dot{\gamma}^2} & 0 & 0\\
0 & \frac{\dot{\gamma}_{\rm anti}^2}{\dot{\gamma}^2}  & \frac{\dot{\gamma}_{\rm anti} \dot{\gamma}_{\rm in}}{\dot{\gamma}^2} \\
0& \frac{\dot{\gamma}_{\rm anti} \dot{\gamma}_{\rm in}}{\dot{\gamma}^2}  & \frac{\dot{\gamma}_{\rm in}^2}{\dot{\gamma}^2} 
\end{bmatrix}, 
\end{equation}
so the reduced set of invariants is $K_3= {\rm tr} (\textbf{N}^3) = 0$, $K_5= {\bf n}^{\rm c} \cdot {\bf N} {\bf n}^{\rm c} = 0$, and $K_6 = {\bf n}^{\rm c} \cdot {\bf N}^2 {\bf n}^{\rm c} = \dot{\gamma}_{\rm in}^2 / 2 \dot{\gamma}^2$. Therefore, the invariants $K_3$ and $K_5$ are always zero in this steady, viscometric flow scenario and cannot differentiate between the in-plane and anti-plane modes of diffusion, so we neglect the dependence of the scalar function $\hat{C}_{\rm diff}$ on $K_3$ and $K_5$. The remaining invariant $K_6$ is bounded within the range $K_6\in [0,1/2]$. In the in-plane mode of diffusion, $\dot\gamma_{\rm anti}=0$, $\dot\gamma = |\dot\gamma_{\rm in}|$, and $K_6=1/2$, and in the anti-plane mode of diffusion, $\dot\gamma_{\rm in}=0$, $\dot\gamma = |\dot\gamma_{\rm anti}|$, and $K_6=0$, so $K_6$ is a suitable invariant for capturing the mode-dependence of the diffusion flux. Renaming this invariant $ K_6 = {\bf n}^{\rm c}\cdot{\bf N}^2{\bf n}^{\rm c} = K^{\rm c}$, the flux equation \eqref{eq:rep_thm} may be written as 
\begin{equation}\label{eq:diff_flux_final}
w^{\rm diff}_i = -\hat{C}_{\rm diff}(I,K^{\rm c}) \bar{d}^2 \dot{\gamma} \frac{\partial c^{\rm l}}{\partial x_i},
\end{equation}
and when a simple linear dependence of $\hat{C}_{\rm diff}$ on $K^{\rm c}$ is adopted, we recover \eqref{eq:gen_diff_coeff} from Section~\ref{sec:3D_constitutive_form}. 

Following an analogous set of arguments, we denote the unit vector along the pressure gradient as ${\bf n}^{\rm P} = (\partial P/\partial{\bf x})/|\partial P/\partial{\bf x}|$ and introduce the combined scalar invariant $K^{\rm P} = {\bf n}^{\rm P}\cdot{\bf N}^2{\bf n}^{\rm P}$. We then write the generalized form for the pressure-gradient-driven segregation flux as follows: 
\begin{equation}\label{eq:seg_flux_final}
w^{\rm P}_i = -\hat{C}^{\rm P}_{\rm seg} (I,J,K^{\rm P}) \frac{\bar{d}^2 \dot{\gamma}}{P} c^{\rm l}(1- c^{\rm l})(1 -  \alpha + \alpha c^{\rm l}) \frac{\partial P}{\partial x_i}, 
\end{equation}
where $J = P/(\bar{d} |\partial P/\partial {\bf x}|)$. In the in-plane mode of segregation, the scalar invariant is $K^{\rm P}=1/2$, and $K^{\rm P} =0$ in the anti-plane mode of segregation. Adopting a linear dependence of $\hat{C}^{\rm P}_{\rm seg}$ on $K^{\rm P}$ gives \eqref{eq:gen_seg_coeff} from Section~\ref{sec:3D_constitutive_form}.

\end{document}